\documentclass[a4paper, traditabstract]{aa}   
\usepackage{graphicx}
\usepackage{lipsum}

\usepackage{comment}
\usepackage{txfonts}
\usepackage{cancel}
\usepackage{float}
\usepackage{multirow}
\usepackage[switch]{lineno}

\usepackage{natbib,twoopt}
\usepackage[breaklinks=true]{hyperref} 
\hypersetup{
  colorlinks   = true, 
  urlcolor     = blue, 
  linkcolor    = blue, 
  citecolor    = blue, 
  breaklinks   = true 
}
\usepackage{bm}
\bibpunct{(}{)}{;}{a}{}{,}             
\usepackage[dvipsnames]{xcolor}

\usepackage{dsfont}
\newcommand{\eqreff}{Eq.~\eqref}

\newcommand{\tabreff}{Table~\ref}
\usepackage{xspace}
\newcommand{\capish}{\texttt{Capish}\xspace}
\begin{document} 

\title{Simulation-based cosmological inference \\ from optically selected galaxy clusters with \texttt{Capish}}

   \author{Constantin Payerne\inst{1}, Calum Murray
          \inst{1,2}, Hugo Simon\inst{1}}
\institute{
Université Paris-Saclay, CEA, IRFU, 91191 Gif-sur-Yvette, France
\and 
Université Paris Cité, CNRS-IN2P3, APC, 75013 Paris, France
}
 \titlerunning{SBI for galaxy cluster cosmology with \capish}
    \authorrunning{C. Payerne et al.}

   \date{Received September 15, 1996; accepted March 16, 1997}
 
  \abstract
  {Galaxy clusters are powerful probes of the growth of cosmic structure through measurements of their abundance as a function of mass and redshift. Extracting precise cosmological constraints from cluster surveys is challenging, as we must contend with nontrivial correlations between lensing mass and optical richness, as well as the complex relationship between richness and the underlying halo mass. These difficulties are compounded by systematic effects such as selection function biases, super-sample covariance, and correlated measurement noise between mass proxies. As upcoming photometric surveys are expected to detect tens to hundreds of thousands of galaxy clusters, controlling these systematics becomes essential. In this paper, we present a forward-modeling approach using simulation-based inference (SBI), which provides a natural framework for jointly modeling cluster abundance and lensing mass observables while capturing systematic uncertainties at higher fidelity than analytic likelihood methods — which rely on simplifying assumptions such as fixed covariances and Gaussianity — without requiring an explicit likelihood formulation. We introduce \texttt{Capish}, a Python code for generating forward-modeled galaxy cluster catalogs using halo mass functions and incorporating observational effects. We perform SBI using neural density estimation with normalizing flows, trained on abundance and mean lensing mass measurements in observed redshift–richness bins. Key cluster-related summary statistics measured on \texttt{Capish} simulations faithfully reproduce their corresponding analytical predictions, and we perform several Bayesian robustness tests of posterior modeling. Our forward model accounts for realistic noise, redshift uncertainties, selection functions, and correlated scatter between lensing mass and observed richness. We find good agreement with explicit-likelihood analyzes, with broader SBI posteriors reflecting the increased realism of the forward model. We also test \texttt{Capish} on cluster catalogs built from a large cosmological simulation, finding a good fit to the cosmological parameters.}
   \maketitle

\begin{figure*}
    \centering
\includegraphics[width=0.99\linewidth]{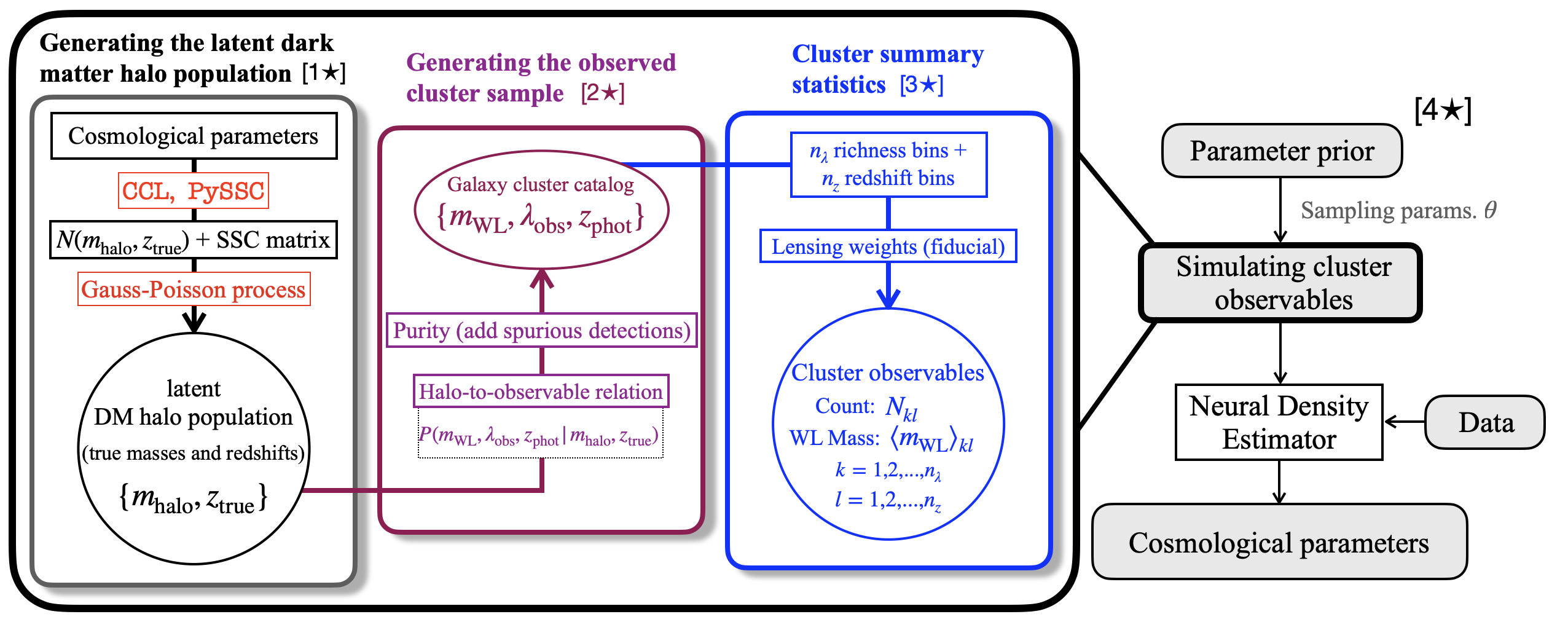}
    \caption{Organization of \capish, detailed in Sect. \ref{sec:capish}. The code is organized in three separate blocks: The first block [1$\star$] is dedicated to generate halo masses and true redshift from an underlying halo mass function, accounting for the effect of SSC (presented in Sect. \ref{sec:capish_dm_halos}) The second block [2$\star$] is dedicated to compute the observed richness and lensing mass for halos, along with photometric redshift (presented in Sect. \ref{sec:capish_galaxy_cluster_cat}). The third block [3$\star$] computes the summary statistics (presented in Sect. \ref{sec:summary_stats}). [4$\star$] denotes the overall SBI structure.}
    \label{fig:capish_scheme}
\end{figure*}

\section{Introduction}

Galaxy clusters form through the gravitational collapse of large matter density fluctuations and emerge as the largest gravitationally bound objects at the intersection of the cosmic web filaments. Therefore, their formation history, mass, and spatial distribution are highly connected to the fluctuations of the matter density field, the expansion rate of the Universe, and the nature of gravity (e.g., \citealt{Bartlett1997,Allen2011clusterreview,Kravtsov2012clusterreview}). 

Constraining cosmological parameters through the abundance of galaxy clusters relies on connecting the observed cluster count to the underlying mass and redshift distribution of the most massive dark matter halos. Cluster masses are not directly observable. With optical surveys, clusters are detected through the density of their member galaxies \citep{Rykoff2014redmapper}, and each cluster is assigned a richness, which is linked to the number of the cluster's galaxy members. The constraining power of cluster number counts is currently limited by our understanding of the cluster scaling relations \citep{Pratt2019massrichness}, linking the underlying cluster mass with what we observe.

Weak gravitational lensing \citep{Bartelmann2001WL} has become a robust tool for constraining cluster masses \citep[e.g.,][]{McClintock2019masscalibration,Umetsu2020clusterlensing, Murray2022lensingmasses,Mistele2024lensing,Grandis2024lensing,Murray2025a}, through the coherent distortion of the shapes of background galaxy images, caused by the bending of the light path due to the cluster's gravitational field.  When available, cluster lensing information has been crucial in tightening the constraints on the cluster mass-proxy relation (see, e.g., \citealt{vonderLinden2014planckclmass,PennaLima2017planckclmass}). 

Over the past two decades, the measurement of the abundance of galaxy clusters has provided competitive constraints on $\Omega_m$ and $\sigma_8$, from optically detected clusters by the Dark Energy Survey\footnote{ Although it should be noted that the \cite{Abbott2020DESCL} found a significant tension with the other DES analysis.} (DES; \citealt{Abbott2020DESCL,Abbott2025DESCLY3,To2021DEScomb,Costanzi2021DESSPTCL}), the Kilo Degree Survey (KiDS; \citealt{Lesci2022KIDSCL,Lesci2025kidscl}), the  Sloan Digital Sky Survey (SDSS; \citealt{Fumagalli2023SDSS,Park2023lensingabundance,Sunayama2023HSClensing}), from X-ray-detected clusters by ROSAT \citep{Mantz2014WTGCL}, eROSITA \citep{Ghirardini2024erositaCL} and from clusters detected through the Sunyaev-Zeldovich (SZ) effect at millimeter wavelengths by the \textit{Planck} satellite \citep{Ade2016PlanckCL,Zubeldia2019PlanckCMBlensing,Salvati2022PlanckSPT,Lee2024PlanckACTCL,Aymerich2024PlanckChandraCL}, the South Pole Telescope (SPT; \citealt{Chaubal2022SPTCLCMBlensing,Bocquet2024SPT,Bocquet2025SPTDES}) or the Atacama Cosmology Telescope (ACT; \citealt{Sehgal2011actCL,Hasselfield2013ACT}). Large cluster surveys such as the Legacy Survey of Space and Time of the Vera Rubin Observatory \citep{LSST} and the \textit{Euclid} mission \citep{laureijs2011euclid} will detect tens to hundreds of thousands of clusters and provide even stronger cosmological constraints. 
The cosmological constraining power of number counts of optically selected clusters is limited by a range of systematic effects that must be understood and controlled. On the theoretical side, these include uncertainties in the modeling of the dark matter density profile in the cluster field \citep{Becker2011modeling,Lee2018wlmass} and in the calibration of the halo mass function \citep{Artis2021hmfcalibration}. Cluster number counts also deviate from a simple Poisson process \citep{Poisson1837}, as fluctuations in the underlying matter density field introduce additional correlated scatter known as super-sample covariance (SSC; \citealt{2003_SV_HU,Gouyou2022SSC}), which constitutes an important source of variance for cluster count measurements \citep{Fumagalli2021pinocchiovariance,Payerne2023,Payerne2024unbinnedSSC}. On the observational side, further systematics arise from the calibration of source galaxy shapes \citep{Hernandez2020shape}, photometric redshift uncertainties in the background galaxy sample \citep{Wright2020photoz}, contamination of the source sample by cluster member galaxies \citep{Varga2019contamination}, and miscentering — offsets between the detected cluster center and the true dark matter halo center (\citealt{Becker2011modeling,Lee2018wlmass,Zhang2019redmappermiscentering,Sommer2022miscentering}, Murray et al., in prep.). Additionally, selection effects from the cluster detection process can introduce correlations between lensing mass and observed richness that can significantly bias cluster abundance and lensing analyzes \citep{Wu2022selection,Zhang2024propertycov,Salcedo2024sbiDESplanck}.

As the volume and precision of cosmological datasets improve, simplified likelihood assumptions tend to break down\footnote{The underlying likelihood function is unknown, unavailable, or too expensive computationally.}, potentially compromising the robustness of inference. Working with next-generation survey data, which are not only high-volume but also high-precision, requires finer resolutions in parameter space and more nuanced modeling of systematics. To address these limitations, the community is increasingly turning to simulation-based inference (SBI), also known as implicit-likelihood inference. From a methodological standpoint, SBI relies on (i) the generation of a synthetic dataset spanning a relevant range of cosmological and nuisance parameters and (ii) learning the probabilistic relationship between the simulated observables and the latent parameters. 

Simulation-based inference has already been successfully applied to real datasets, for example, DES \citep{Jeffrey2021sbi,Gatti2025dessbi,Prat2025dessbi}, KiDS \citep{Kramsta2025kidssbi}, SDSS \citep{Lemos2023simbigsdss}, and \textit{Planck} cluster number counts \citep{Zubeldia2025sbicl}, showing good agreement with explicit-likelihood approaches while naturally accounting for stochastic effects beyond the limitations of Gaussian likelihood assumptions.
Several proof-of-concept studies, validated on mock datasets \citep{Reza2022sbi,Reza2024sbi,Kosiba2025sbiCL,Cerardi2025Xray,Regamey2025sbicl,Saez2026clusterCNN}, have explored the potential of SBI from X-ray, SZ, and optical cluster samples, paving the way for the broader adoption of likelihood-free inference techniques in the analysis of current and upcoming cluster surveys.

In this paper, we present \capish\footnote{\url{https://github.com/calumhrmurray/capish}} (Cluster abundance posterior inference from simulated halos), a Python code dedicated to performing the cosmological analysis of optically selected cluster catalogs in the era of the Vera C. Rubin LSST and the \textit{Euclid} mission, using simulations. In Sect. \ref{sec:capish} we present the different functionalities of the \capish code. Its diagram is shown in Fig. \ref{fig:capish_scheme}. We evaluate in Sect. \ref{sec:capish_validation} the internal calibration of \capish simulations and self-applied cosmological analyzes, as well as by using the \textit{Euclid}
Flagship dark matter halo catalog. We conclude in Sect. \ref{sec:conclusions}. 

\section{The \capish code}
\label{sec:capish}

\capish is dedicated to (i) simulate analytically galaxy cluster catalogs, given cosmological parameters, survey characteristics, and some observational systematics, (ii) training neural density estimators (NDE), and (iii) performing the cosmological analyzes of cluster catalogs from the trained NDE. The code structure of \capish is presented in Fig. \ref{fig:capish_scheme}.

\subsection{Generating dark matter halo catalogs}
\label{sec:capish_dm_halos}
\capish generates a dark matter halo catalog from a halo mass function (e.g., \citealt{Tinker2008hmf,Tinker_2010,Despali_2015}). 
We first compute the halo count prediction $N_{ij}$ in narrow mass bins (between $\log_{10}m=12$ to $\log_{10}m=16.5$)  with width $\Delta \log_{10}m_i \ll 1$ and redshift bins (between $z=0$ to $z=1.2$) with width $\Delta z_j \ll 1$, given as
\begin{equation}
    N_{ij}^{\rm grid} =  \Delta \log_{10}(m_i)\, \Delta z_j \ n_h\left(\log_{10}m_i^{\rm center},z_j^{\rm center}\right),  
    \label{eq:N_count_halo_th}
\end{equation}
where $\log_{10}m_i^{\rm center}$ (respectively, $z_j^{\rm center}$) is the center of the $i$-th log-mass (respectively, $j$-th redshift) bin. In the above equation, $n_h(m,z)$ is the predicted total halo number density per mass and redshift range given by
\begin{equation}
n_h(\log_{10}m,z)=\Omega_S\,\frac{dn(\log_{10}m, z)}{d\log_{10}m}\,\frac{d^2V(z)}{dz\, d\Omega},
    \label{eq:dn_dm_dz}
\end{equation}
where $dn(m,z)/dm$ is the halo mass function of objects at redshift $z$ and mass $m$, $\Omega_S$ is the survey solid angle and $V(z)$ the comoving volume\footnote{For the computation of both the halo mass function and comoving volume, we use the LSST DESC Core Cosmology Library \citep{Chrasi2019ccl}.}. 

The intrinsic clustering of the underlying matter density field induces a covariance in halo counts. This is often referred to as SSC (\citealt{2003_SV_HU}), and is an important contribution to the variance of large-scale structure probes (particularly galaxy clusters, \citealt{Fumagalli2021pinocchiovariance,Payerne2023}) as cosmological surveys increase in coverage and depth. The SSC is usually accounted for in the binned count analytic likelihood as an additional covariance term. The effect of SSC is as a correlated scattering of the per-bin halo count prediction in \eqreff{eq:N_count_halo_th}, such as \begin{equation}
    \widetilde{N}_{ij}^{\rm grid}= N_{ij}^{\rm grid}\, (1+\delta_{ij}^{\rm SSC})
    \label{eq:Nij_tilde}
,\end{equation} where $\delta_{ij}^{\rm SSC}$ is the scattering associated with the SSC, which satisfies $\langle\delta_{ij}^{\rm SSC}\rangle = 0$ and $\langle\delta_{ij}^{\rm SSC}\delta_{kl}^{\rm SSC}\rangle = b_{ij}\, b_{kl}\, \sigma_{{\rm SSC},jl}^2$, where $b_{ij}$ is the halo bias at mass $m_i$ and redshift $z_j$, and \citep{Lacasa2018sscapprox,Lacasa2023SSC}
\begin{equation}
    \sigma_{{\rm SSC},jl}^2 = \frac{4\pi}{\Omega_S}\int \frac{k^2 dk}{2\pi^2}\, j_0[kw(z_j)]\, j_0[kw(z_l)]\, P_{\rm mm}(k|z_j, z_l)
\label{eq:sigma2_z1z2_fullsky}
\end{equation}
is the (partial-sky) amplitude of matter density fluctuations\footnote{We use \texttt{PySSC} \citep{Lacasa19,Gouyou2022SSC} to compute $\sigma_{{\rm SSC},jl}$ (we use the \texttt{PySSC} function \texttt{sigma2}$\_$\texttt{fullsky}, then rescaled by the sky fraction $\Omega_S/4\pi$). The code is available at \url{https://github.com/fabienlacasa/PySSC}.}. 
In the equation above, $j_0(x) = \sin(x)/x$ is the zero-th order spherical Bessel function, and $P_{\rm mm}(k|z_1, z_2)$ is the linear matter power spectrum. 
In the unbinned regime (given by $\Delta z_j \ll 1$; see, e.g., \citealt{Mantz2010CCmethodunbinned,2014_PennaLima}), the variance of $\delta_{ij}^{\rm SSC}$ becomes
maximal, such as the occurrence $\delta_{ij}^{\rm SSC} < -1$ becomes statistically significant if $\delta_{ij}^{\rm SSC}$ is considered to be a Gaussian variable. This could yield to unphysical halo counts through \eqreff{eq:Nij_tilde} \citep{Payerne2024unbinnedSSC}. In this regime, it is more appropriate to consider $\delta_{ij}^{\rm SSC}$ as a log-normal random variable \citep{Coles1991density,Wen2020unbinnedPNL}, such as $\ln(1+\delta^{\rm SSC})\sim \mathcal{N}(\mu^{\rm SSC}, \Sigma^{\rm SSC})$ where
\begin{align}
    \mu^{\rm SSC}_{ij} &= - \frac{1}{2}\ln(1 + b_{ij}^2\, \sigma_{{\rm SSC},jj}^2),\\ \Sigma^{\rm SSC}_{ij,kl} &= \ln(1 + b_{ij}\, b_{kl}\, \sigma_{{\rm SSC},jl}^2),
    \label{eq:mu_sigma_logprob_ssc}
\end{align}
such that the aforementioned conditions for $\langle\delta_{ij}^{\rm SSC}\rangle$ and $\langle\delta_{ij}^{\rm SSC}\delta_{kl}^{\rm SSC}\rangle$ are satisfied. Finally, the number of halos is sampled from a Poisson distribution $\mathcal{P}$ such as
\begin{equation}
     \widehat{N}^{\rm grid}_{ij}\sim \mathcal{P}[\widetilde{N}^{\rm grid}_{ij}].\\
\end{equation}
The resulting dark matter halo catalog is obtained by collecting $\widehat{N}_{ij}^{\rm grid}$ times samples drawn randomly inside the mass-redshift bin, with centers $m_i$ and $z_j$. 

In the upper panel of  Fig. \ref{fig:bias_mass_redshift}, we show the ratio between (i) the mean of 200 simulated dark matter halo counts with \capish in large mass-redshift bins (for display purposes, to compare with binned prediction) and (ii) the associated theoretical prediction $N_{ij}^{\rm true}$, computed as the two-dimensional integral of the halo density in \eqreff{eq:dn_dm_dz} within the same large mass-redshift bins.
The ratio is close to one for most mass scales, except at high mass, where the measured halo abundance is smaller than the cosmological prediction because of the finite volume of the simulated catalog. Additionally, we compute the dispersion of halo number counts in the same mass-redshift bins, which we compare to the theoretical variance given by\footnote{We use the \citet{Tinker_2010} halo bias for the SSC part.} \citep{2014_Takada,Lacasa19}
\begin{equation}
    \mathrm{Var}(N_{ij}) = N_{ij}^{\rm true} + (N_{ij}^{\rm true})^2\langle\, b_{ij}\,\rangle^2 \, S_{jj}
    \label{eq:var_sn_ssc_true}
\end{equation}
as the sum of a Poisson term and an SSC term, where $\langle b_{ij}\rangle$ is the mean halo bias in the mass-redshift bin and $S_{kl}$ is given by\footnote{We use the \texttt{Sij\_fullsky} function of \texttt{PySSC}, then rescaled to the sky fraction $\Omega_S/4\pi$.} \citep{Lacasa2018sscapprox} 
\begin{equation}
     S_{kl} = \int_{z_{k}}^{z_{k+1}}\int_{z_{l}}^{z_{l+1}} \,\frac{dV(z')\, dV(z'')}{V_k\,V_l}\,\sigma^2_{{\rm SSC}}(z', z''),
     \label{eq:S_ij}
\end{equation}
where $dV(z)$ is the comoving volume per steradian ($V_k$ is the total volume in the $k$-th redshift bin). In the bottom panel of  Fig. \ref{fig:bias_mass_redshift}, we see that the measured count dispersion has a good agreement with analytic count variance at most mass scales. We also see that SSC is important for lower mass halos; therefore, from this example, we show that our implementation of SSC in our simulation-based framework is in agreement with the behavior of SSC in the analytical likelihood analysis.  The resulting dark matter halo catalog serves as a basis to generate the catalog of observed galaxy clusters.
\subsection{Generating the galaxy cluster catalog}
\label{sec:capish_galaxy_cluster_cat}
\subsubsection{From halo properties to cluster observables}
The dark matter halo catalogs have to be connected to observational cluster properties, such as the observed richness and the weak gravitational lensing signal. We use a general multi-variate Gaussian model \citep{Evrard2014massobservable,Zhang2024propertycov,Payerne2025cosmodc2} for the observed richness $\lambda_{\rm obs}$ and a lensing observable, corresponding to the lensing mass $\log_{10}m_{\rm WL}$, i.e.,
\begin{align}
\label{eq:plambda}
&\ln\lambda_{\rm obs}, \log_{10}m_{\rm WL}  \sim \mathcal{N}[\mathrm{mean}(m,z), \mathrm{Cov}(m,z)]
\end{align}
where \begin{equation}
    \mathrm{mean}(m,z) = 
\{
\langle \ln \lambda_{\rm obs}|m,z\rangle,\langle \log_{10}m_{\rm WL}|m,z\rangle\}
\end{equation} and
\begin{equation}
\mathrm{Cov}(m,z)= 
\begin{pmatrix}
\sigma_{\ln\lambda}^2 & \rho\sigma_{\ln \lambda}\sigma_{\rm WL}   \\
  \rho\sigma_{\ln \lambda}\sigma_{\rm WL} & \sigma_{\rm WL}^2 
\label{eq:covariance_plambda}
\end{pmatrix}.
\end{equation}

We define the cluster richness $\lambda_{\rm obs}$  as related to the count of cluster's member galaxies -- intrinsically linked to the cluster's formation history -- evaluated, for example, within a circular aperture centered on the detected cluster position (à la \texttt{redMaPPer}, \citealt{Rykoff2014redmapper}). The observed richness is linked to the specific selection of these galaxies in photometric surveys; it depends on (i) the cluster detection algorithm to define and measure the cluster's richness\footnote{Generally scaling with, but not equal to, the number counts of cluster member galaxies. For instance, \citet{Rykoff2014redmapper} defines the richness as the sum of membership probabilities for red-sequence selected galaxies in the vicinity of a cluster, where \citet{Aguena2021WAZP} defines clusters as the overlap of multiple galaxies' photometric redshift distributions happening at the same position on the sky.} \citep{Rykoff2014redmapper,Bellagamba2017AMICO,Aguena2021WAZP} (ii) observational and/or detection noise.   A power-law relation \citep[see, e.g.,][]{Mantz2008cluster,Evrard2014massobservable,Saro2015massrichness,Farahi2018cov,Murata2019HSCrichnessmassrelation,Anbajagane2020stellat} is commonly adopted to connect the mean cluster's observed (log-) richness $\ln \lambda_{\rm obs}$ with its halo mass and redshift, via
\begin{equation}
    \langle \ln \lambda_{\rm obs}|m,z\rangle = \mu_0^{\lambda} + \mu_m^{\lambda} \log_{10} \left(\frac{m}{m_0}\right) + \mu_z^{\lambda}\log\left(\frac{1+z}{1+z_0}\right).
\label{eq:powerlaw_richness}
\end{equation}
Two contributions affect the measurement scatter of the cluster's richness. First, the intrinsic scatter ($\sigma_{\ln \lambda, \rm int}$) denotes the contribution of the intrinsic cluster's formation history. Second, Poisson sampling arises as a statistical process on its own, associated with the observation, since richness is a "count-in-cell" observable. In this work, we use
\citep{Farahi2018cov,Zhang2023triax,To2025desy6}
\begin{equation}    \sigma_{\ln \lambda}^2 = \sigma_{\ln \lambda, \rm int}^2 + \exp\{-\langle \ln \lambda_{\rm obs}|m,z\rangle\},
    \label{eq:sigma2_lnlambda}
\end{equation}
where the first term denotes the intrinsic scatter and the second the Poisson dispersion\footnote{
Most papers use $(\exp(\langle \ln \lambda_{\rm obs}|m,z\rangle)-1)/\exp(2\langle \ln \lambda_{\rm obs}|m,z\rangle)$, which is a good approximation for large richness values. Since we want our code to be stable over the full richness range, we use \eqreff{eq:sigma2_lnlambda}.} of the richness measurement. This is currently the primary weakness of \capish, as the actual relation between the halo mass and the observed cluster richness is likely much more complex.

Galaxy cluster lensing masses are primarily inferred from the weak gravitational lensing effects of the cluster's gravitational potential, inducing a small distortion in the shapes of background galaxies. This is generally estimated from the excess surface density \citep{Murray2022lensingmasses}, which is taken to be the average in the projected radial bins around the cluster center of the weighted background ellipticities $
\widehat{\Delta\Sigma}(R)=\langle \Sigma_{\rm crit}(z_s,z_l)\,\epsilon_+^{l,s}\rangle(R)$ where $\Sigma_{\rm crit}(z_s,z_l)$ is the critical surface mass density -- a geometrical lensing factor -- and $\epsilon_+^{s}$ is the tangential ellipticity of a source at $z_s>z_{\rm cl}$ within the radial bin. The excess surface density in radial bins is not the unique cluster weak lensing signal estimator that is used in the literature; one can also use binned reduced shear \citep{Becker2011modeling}, two-dimensional lensing maps (\citealt{Oguri2010ellipticity}, Murray et al., in prep.), or weak lensing shear magnification \citep{Murray2025a}. The measured excess surface density profile probes $\Sigma(R|m)$ are the projected three-dimensional matter density $\rho_{\rm cl}(\vec{x})$ in the cluster field.
The cluster lensing mass $m_{\rm WL}$ (along with other halo parameters such as its concentration) is obtained by fitting the lensing profile with the appropriate model for the matter density around the cluster center\footnote{We generally consider that the lensing signal is dominated by the 1-halo term at scales below 3 Mpc, originating from the cluster itself, and the 2-halo term becomes increasingly important at larger scales, originating from neighboring halos.} \citep{McClintock2019masscalibration}. The lensing mass $m\rightarrow m_{\rm WL}$ could be biased with respect to its true underlying halo mass, so we use the power-law relation 
\begin{equation}
    \langle \log_{10}m_{\rm WL}|m,z\rangle  = \mu^{\rm WL}_0 + \mu^{\rm WL}_m\log_{10}m + \mu_z^{\rm WL}\ln\left(\frac{1+z}{1+z_0}\right).
    \label{eq:log10_mwl_mean}
\end{equation}
In the above equation, any deviation from $\mu^{\rm WL}_0=0$, $\mu^{\rm WL}_m=1$, and $\mu_z^{\rm WL}=0$ indicates that the lensing masses are biased with respect to the true halo mass, which may happen if the underlying dark matter density profile is not accurately representing lensing data. 

The lensing profile $\widehat{\Delta\Sigma}(R)$ is a scattered measurement, arising from several independent contributions  \citep{Hoekstra2001lss,Hoekstra2003lensing,Becker2011modeling,2015MNRAS.449.4264G,Wu2019covarianceDeltaSigma}. Its dominant noise term ($\sigma_{\rm WLgal}^{\rm \Delta\Sigma}$) arises from intrinsic galaxy shape dispersion and limited source galaxy samples for the lensing profile estimation, scaling as $\sigma_\epsilon/\sqrt{\bar{n}_{\mathrm{gal}}(z)}$, where $\sigma_\epsilon^2=\sigma_{\epsilon, \mathrm{meas.}}^2+\sigma_{\epsilon, \mathrm{SN}}^2$ being the dispersion of galaxy shapes due to (i) intrinsic galaxy shape noise ($\sigma_{\epsilon, \mathrm{SN}}\approx 0.25$, \citealt{Chang2013density}), (ii) errors in shape measurement algorithms ($\sigma_{\epsilon, \mathrm{meas.}}$), and $\bar{n}_{\mathrm{gal}}(z)$ the surface number density of lensed background galaxies for a cluster at redshift $z$. Additional noise—independent of source density—comes from the intrinsic dispersion in halo morphology\footnote{Because most of the time we use a simplified spherical model to constrain the mass of nonspherical clusters.} ($\sigma_{\rm WLint}^{\rm \Delta\Sigma}$) and the impact of correlated and/or uncorrelated large-scale structure ($\sigma_{\rm WLcLSS}^{\rm \Delta\Sigma}$ and $\sigma_{\rm WLuLSS}^{\rm \Delta\Sigma}$). On the observational side, further systematics arise from shape measurement errors \citep{Hernandez2020shape}, photometric-redshift uncertainties \citep{Wright2020photoz}, cluster-member contamination \citep{Varga2019contamination}, and miscentering \citep{Zhang2019redmappermiscentering,Sommer2022miscentering}. In summary, the total variance $\sigma_{\rm WL}^{\rm \Delta\Sigma}$ of the measured excess surface density profile $\Delta\Sigma(R)$ is given by the quadratic sum of these terms. The lensing profile variance depends on the underlying halo mass, its redshift, the cluster's local environment, and finally, the lensing survey characteristics. In this work, we rely on the lensing mass; its variance -- as for the lensing profile -- can be decomposed as
\begin{equation}
    \sigma_{\rm WL}^2 = \sigma_{\rm WL gal}^2+\sigma_{\rm WLint}^2 +\sigma_{\rm WLcLSS}^2+\sigma_{\rm WLuLSS}^2.
\label{eq:sigma_WL}
\end{equation}
The above depends on the number of parameters (and correlations) that are fitted jointly with the lensing mass (see Appendix \ref{app:error_model_leisng_mass}). For simplicity, in this work we only consider the $\sigma_{\rm WL gal}^2$ component, and leave the implementation of the other terms for future works, which can be obtained separately from the theory \citep{Shirasaki2018covariancelensing} of dedicated simulations (\citealt{Becker2011modeling,Wu2019covarianceDeltaSigma}, Murray et al. in prep.). As the number density of sources approaches that expected from space-based observations or LSST conditions (e.g., DES provided $\bar{n}_{\mathrm{gal}} = 10$ arcmin$^{-2}$, where LSST will provide $\bar{n}_{\mathrm{gal}} = 40$ arcmin$^{-2}$), the contribution to the scatter in the weak lensing masses from galaxy shape noise becomes comparable or even subdominant to the intrinsic scatter terms and large-scale structures \citep{Hoekstra2001lss,Becker2011modeling}. A fixed parameter can be used, but \capish also provides an information matrix-based analytical model $\sigma_{\rm WL gal}(m,z)$ as explained in Appendix \ref{app:error_model_leisng_mass}, for improved realism, as shown in Fig. \ref{fig:sigmaWL}, accounting for (i) the cluster's redshift-dependent number density of source galaxies, (ii) the theoretical modeling for $\Delta\Sigma(R)$, and (iii) the considered radial fitting range. In this error model, the scatter is larger for low-mass halos (whose weak lensing signals have a lower signal-to-noise ratio) or high-redshift halos (which have fewer source galaxies to measure their lensing profiles).

Within \capish, it is also possible instead to add lensing mass scatter at the summary statistics level (the detailed description of the summary statistics we used in this study -- evaluated in bins of richness and redshift -- is discussed later in Sect. \ref{sec:summary_stats}). First, for each $i$-th halo in the halo catalog, we computed its lensing mass  $\log_{10}m_{\mathrm{WL},i}=\langle \log_{10}m_{\rm WL}|m_i,z_i\rangle$ by using \eqreff{eq:log10_mwl_mean} (i.e., no individual lensing mass dispersion). Then, for an ensemble (or stack) of $N_{\rm cl}$ galaxy clusters within a given richness-redshift bin, the mean halo mass within the stack is scattered accordingly with $\sigma_{\rm WL}^{\rm stack}=\sigma_{\rm WL}/\sqrt{N_{\rm cl}}$. This is closer to the modeling used in recent cluster cosmology analyzes and weak lensing mass-relation calibration techniques \citep{Abbott2020DESCL,Lesci2025kidscl,Payerne2025cosmodc2}, where the mean mass of clusters is inferred from a clusters' stacked weak lensing profiles\footnote{In this case, there is no difference in considering $\log_{10}m_{\rm WL}$ or $m_{\rm WL}$ as the random Gaussian lensing variable.}.

For optically selected clusters, an important systematic is the covariance between the lensing observable and the observed richness that is measured by cluster finders. This correlation coefficient is effective, with no distinction between intrinsic and extrinsic scatters \citep{Wu2019covarianceDeltaSigma,Zhou2024selectionbias}; the intrinsic part of this covariance reflects galaxy assembly bias tied to secondary halo properties such as concentration, mass accretion rate, or dynamical state \citep{Zhang2024propertycov}. The extrinsic part arises from selection biases within cluster-finding algorithms, shaped by observational systematics, including photometric-redshift errors \citep{Graham2017photoz}, shear shape noise \citep{Wu2019covarianceDeltaSigma}, projection and percolation effects \citep{Costanzi2019projeffects}, triaxiality bias \citep{Zhang2023triax}, and additional complexities associated with realistic cluster finders such as \texttt{redMaPPer} \citep{Rykoff2014redmapper}. This covariance induces additive biases that cannot be mitigated by increasing the source density or reducing shape noise \citep{Nord2008selection,Evrard2014massobservable,Farahi2018cov,Wu2019covarianceDeltaSigma,Nde2025selectioneffect}, but must be accurately quantified to achieve a percent-level mass calibration \citep{Rozo2014scalingrel,Zhou2024selectionbias}. \citet{Salcedo2025desSBI} showed that projection effects that impact weak-lensing measurements of DES~Y1 clusters can, if not properly accounted for, induce significant biases in inferred cosmological parameters \citep{Abbott2020DESCL}. These effects can be efficiently modeled using a simulation-based forward approach to stacked lensing observables.  In \capish, we adopt a single parameter, $\rho = \mathrm{Corr}(\log_{10} m_{\rm WL}, \ln \lambda_{\rm obs} \mid m, z)$, to describe the overall correlation between the lensing observable and the richness observed assigned by the cluster finder at a fixed true mass and redshift. We further allow $\rho$ to vary with mass and redshift, as selection biases—arising from the mechanisms discussed above—are expected to depend on local cluster properties and environment and therefore lead to a mass-dependent correlation between cluster lensing and cluster richness.

\subsubsection{Cluster photometric redshifts}
Individual observed clusters' redshifts are usually derived from their member galaxies (see, e.g., \citealt{Rykoff2014redmapper}). It is linked to the redshift of its underlying dark matter halo through
\begin{equation}
z_{\mathrm{phot}} \sim  \mathcal{N}\bigl(z + b(z), \sigma_{\rm PZ}(z)^2\bigr),
\label{eq:photometric_redshift}
\end{equation}
where $b(z)$ represents a possible systematic bias, and $\mathcal{N}(0, \sigma_{\rm PZ}^2)$ denotes a Gaussian random variable with zero mean and redshift-dependent scatter $\sigma_{\rm PZ}(z) = \sigma_{\rm PZ,0} (1 + z)$, and depends on the optical survey characteristics, depth, and so on, setting $\sigma_{\rm PZ,0}$ at $z=0$. This functional form captures the typical increase in uncertainty at a higher redshift due to observational limitations. It is, however, simplistic (although sufficient for this proof-of-concept study) since in practice cluster redshift estimates are affected by additional systematic effects, including filter transitions of key spectral features used in redshift reconstruction. For example, the 4000\,\AA\ break shifting between the $g$ and $r$ bands at $z \sim 0.4$ can impact both the inferred cluster redshift and its uncertainty (see, e.g., \citealt{Kluge2024erositacl}). Moreover, the uncertainty model depends on the number and wavelength coverage of the available filters. In particular, low-redshift reconstruction ($z < 0.05$) could be significantly improved with LSST $u$-band observations \citep{Ivezic2019lsst}, which extend to shorter wavelengths than the current blue limit set by the $g$ band.

\begin{figure*}
\begin{center}
\includegraphics[width=0.49\textwidth]{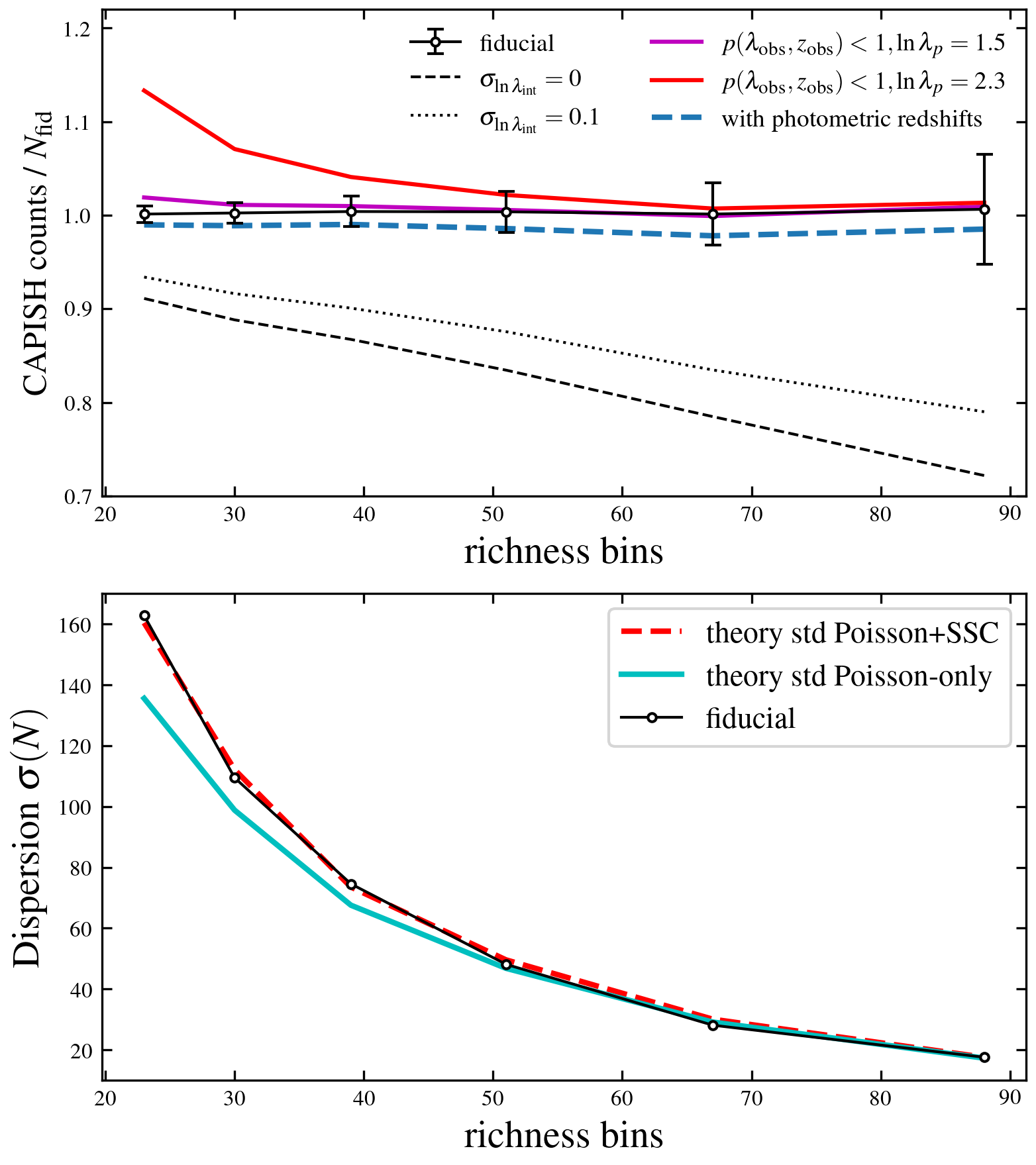}
\includegraphics[width=0.49\textwidth]{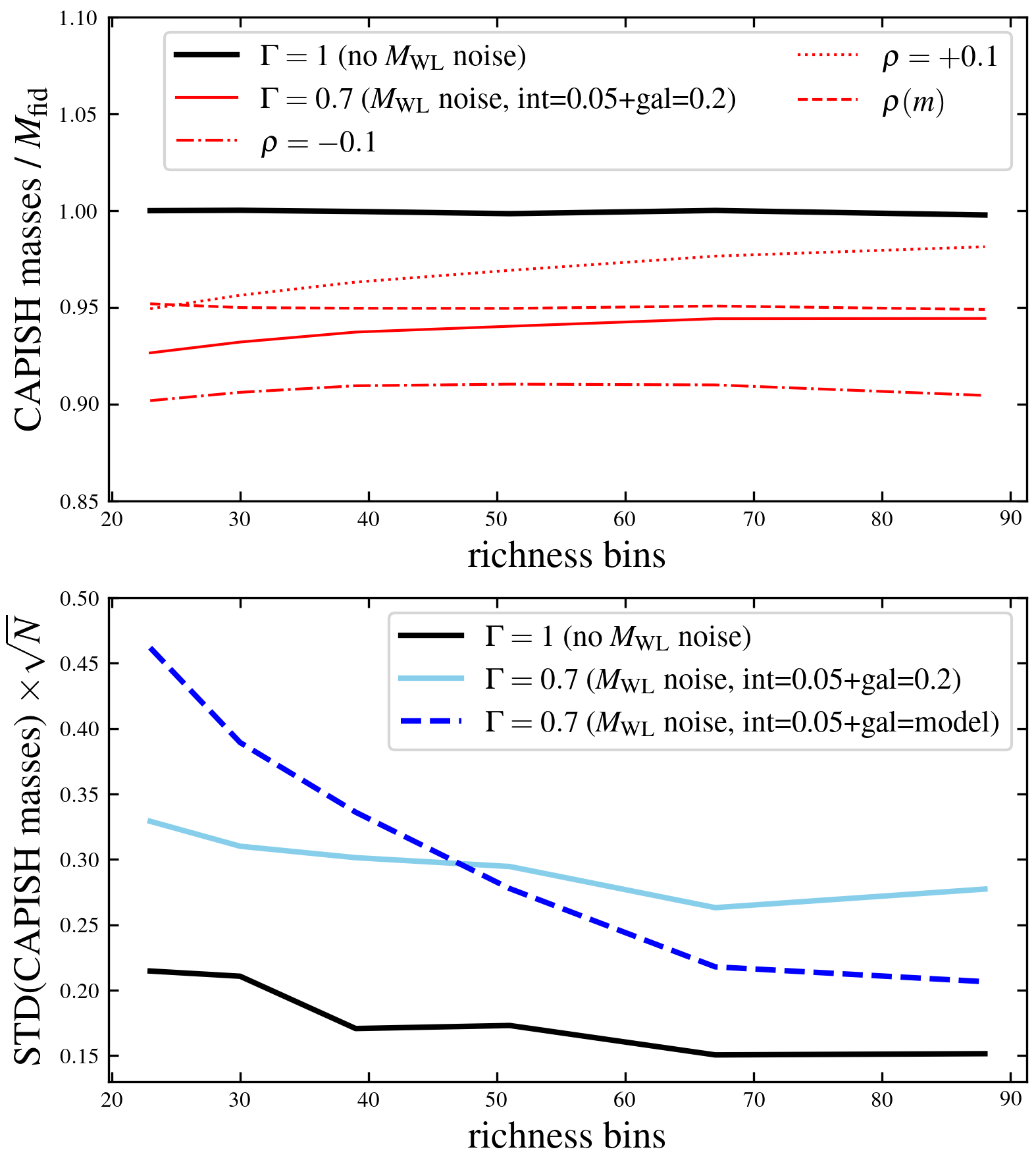}
\caption{\textit{Top left}: \capish mean count outputs, averaged over 100 simulations, compared to the theoretical prediction for the mean cluster count in redshift richness bins. \textit{Bottom left}: Standard deviation of the same counts over the 100 simulations. \textit{Top right}: \capish mean mass outputs, averaged over 100 simulations, compared to the theoretical prediction for the mean cluster masses in redshift-richness bins. \textit{Bottom right}: Standard deviation of the same mean masses over the 100 simulations.}
\label{fig:capish_outputs_mass}
\end{center}
\end{figure*} 

\subsubsection{Cluster selection function}

To account for the cluster selection function arising from the performance of the cluster finder, the ideal framework would forward model the full pipeline: galaxy catalog, cluster detection algorithm, and richness assignment. This is beyond our current scope. We aim, then, to include cluster finding characteristics in the simulated cluster catalog. The first effect in cluster finding is the presence of spurious detection, encoded in the cluster finder purity; optical surveys suffer from artifacts--- unmasked stellar diffraction spikes, globular clusters misidentified as galaxy overdensities, line-of-sight projections---that inject spurious detections into the catalog. We therefore include a purity model in \texttt{Capish} to account for these false positives. We define purity $p(\lambda_{\rm obs}, z_{\rm phot})$ as the fraction of detected objects that correspond to genuine clusters. Following \citet{Aguena2018completenesspurity}, we adopt the form
\begin{equation}
p(\lambda_{\rm obs}, z_{\rm obs}) = \frac{(\lambda_{\rm obs}/\lambda_p)^{n_{\rm pur}(z_{\rm obs})}}{1 + (\lambda_{\rm obs}/\lambda_p)^{n_{\rm pur}(z_{\rm obs})}},
\label{eq:purity}
\end{equation}
where $n_{\rm pur}$ and $\lambda_p$ set the shape and scale of the purity function. This parameterization ensures purity approaches unity at high richness while allowing significant contamination at low richness, consistent with the behavior of real cluster finders. For each richness-redshift bin, we first compute the histogram of true detections $\hat{N}_{\rm true}(\lambda_{k,\rm obs}, z_{l,\rm phot})$, then draw the number of spurious detections from
\begin{equation}
\hat{N}_{kl,\rm fake} \sim \mathcal{P}\left(\hat{N}_{\rm true}(\lambda_{k,\rm obs}, z_{l,\rm phot})\times\frac{1 - p(\lambda_{k,\rm obs}, z_{l,\rm phot})}{p(\lambda_{k,\rm obs}, z_{l,\rm phot})}\right).
\end{equation}
Each spurious detection is assigned a richness and redshift drawn uniformly within its bin. We set the lensing mass of these objects to zero, reflecting the assumption that they produce a negligible weak lensing signal\footnote{Any identified structure would inevitably produce a positive lensing signal, as it may consist of a low-mass population or result from the cumulative contribution of projected structures along the line of sight; however, this has not been explored extensively in the literature, and this is not the purpose of this work, so we keep using the aforementioned assumption.}.

Another feature in the cluster selection function is the cluster catalog completeness, denoting the underlying halo population that is effectively probed by the detected galaxy clusters. Working at the catalog level, imposing a low richness cut (e.g., $\lambda_{\rm obs} >20$) restricts the probed halo masses above a certain value, determined by the richness-mass relation (and its scatter) in \eqreff{eq:powerlaw_richness}. As an illustration, Fig. \ref{fig:completeness} shows how completeness (given by the ratio between the detected clusters that are true halos and the true halo population) varies with halo mass for different richness thresholds and intrinsic scatter values. We see that lower thresholds and smaller scatter both steepen the selection function and push it to lower masses. As a result, the effect of completeness is inherent in the presence of a statistical mass-richness relation of the cluster catalog.

Although not implemented here, Appendix~\ref{app:alternative_completeness} discusses an alternative completeness definition. Following \citet{Rozo2010CLSDSS,Aguena2018completenesspurity,Euclid2019clusterchallenge,Lesci2022KIDSCL,Lesci2025kidscl,Payerne2025cosmodc2}, completeness can be modeled as a mass- and redshift-dependent function $c(m,z)$, independent of the mass--richness relation in \eqreff{eq:powerlaw_richness}. It quantifies the fraction of halos missed by the cluster finder due to intrinsic detection limitations, rather than the effective mass threshold of the richness cut (see Fig. \ref{fig:completeness}). Incorporating $c(m,z)<1$ would reduce the predicted cluster counts, in contrast to purity, which increases them.

\subsection{Cluster count and cluster lensing summary statistics}
\label{sec:summary_stats}
\capish returns two summary statistics: 
\begin{itemize}
    \item Count in observed richness-redshift bins: The cluster abundance summary statistic is given by
    \begin{equation}
        \widehat{N}_{kl} = \sum\limits_{\rm cl}\mathds{1} \{\mathrm{cl} \in (k,l)\},
        \label{eq:count}
    \end{equation}
    which gives the measured counts in bins of observed richness and photometric redshifts. In the above equation, $\mathds{1} \{\mathrm{cl} \in (k,l)\} = 1$ if the cluster belongs to the $(k,l)$ richness--redshift bin, and $0$ otherwise.
    \item Mean mass in observed richness-redshift bins: The mean mass in a given richness-redshift bin is given as the mean of individual lensing masses. However, the usual way to recover the mean lensing mass of clusters within the $kl$-th observed richness-observed redshift bins is to fit the corresponding stacked lensing profile $\widehat{\Delta\Sigma}_{kl}$ in the same $kl$ richness-redshift bin (see, e.g., \citealt{Murata2019HSCrichnessmassrelation}). From this, the link between the "true" mean mass of a stack and the mean mass inferred from the weak lensing stacked profile $\widehat{\Delta\Sigma}_{kl}$ may not be trivial and may be affected by the dispersion of individual masses within the stack \citep{McClintock2019masscalibration}. \citet{Melchior2017logslope} proposed that since $\Delta\Sigma(R|M) \propto M^{\Gamma(R|M)}$ (at first order) where $\Gamma(R|M)$ is the logarithmic slope of the excess surface density\footnote{For instance, \citet{Melchior2017logslope} found a typical value of $\Gamma = 0.74$ for a Navarro–Frenk–White (NFW; \citealt{Navarro1997nfw}) profile within the one-halo regime.}, we use $\Gamma=0.7$ in this work; the mean weak lensing mass inferred from $\widehat{\Delta\Sigma}_{kl}$  \citep{Abbott2020DESCL,Lesci2022KIDSCL,Fumagalli2023SDSS} can be modeled and/or interpreted as 
    \begin{equation}
        \widehat{M}_{kl} = \left(\frac{1}{\sum\limits_{\mathrm{cl}}W(z_{\rm cl})\,\mathds{1} \{\mathrm{cl} \in (k,l)\}}\sum\limits_{\mathrm{cl}}W(z_{\rm cl})\,  m_{\mathrm{cl},\rm WL}^\Gamma\, \mathds{1} \{\mathrm{cl} \in (k,l)\}\right)^{1/\Gamma}\label{eq:stacked_mass_gamma}
    \end{equation} 
    where,considering $\sigma_{\epsilon}^2(z_s) =\sigma_{\epsilon, \mathrm{meas.}}^2(z_s)+\sigma_{\epsilon, \mathrm{SN}}^2(z_s)$ (in this work, we set $\sigma_{\epsilon, \mathrm{meas.}}=0$), the weights $W(z_{\rm cl})$ are the maximum signal-to-noise ratio weights \citep{Sheldon2004DSestimator} and are given by
    \begin{equation}
        W(z_{\rm cl}) = \frac{1}{\int_{z_{\rm cl}}^{\infty} dz_s' \, n_{\rm gal}(z_s')}\int_{z_{\rm cl}}^{\infty} \,dz_s' \, n_{\rm gal}(z_s') \,\frac{\Sigma_{\rm crit}^{-2}(z_s', z_{\rm cl})}{\sigma_{\epsilon}^2(z_s')},
    \end{equation}
    where $n_{\rm gal}(z)$ is the source galaxy redshift distribution, taken from \citet{Chang2013density}.
    
    As stated in Sect. \ref{sec:capish_galaxy_cluster_cat}, a choice is to consider that (i) individual masses are not scattered in \eqreff{eq:covariance_plambda}, and (ii) that the weak lensing mass scatter is instead applied at the stacked level, such as
\begin{equation}
    \log_{10}\widehat{M}_{kl} \rightarrow \log_{10}\widehat{M}_{kl} + \widehat{\epsilon}_{kl},
    \label{eq:sigma_wl_stack}
\end{equation}
where $\widehat{\epsilon}_{kl} \sim \mathcal{N}\left(0, (\sigma_{\mathrm{WL}, kl}^{\rm stack})^2\right)$. The stack scatter is given by
$\sigma_{\mathrm{WL}, kl}^{\rm stack} = \sigma_{\mathrm{WL}, kl} / \sqrt{\widehat{N}_{kl}}$,
with $\sigma_{\mathrm{WL}, kl}$ denoting the weak lensing mass scatter for a single cluster lying in the $(k,l)$ richness-redshift bin. This quantity $\sigma_{\mathrm{WL}, kl}$ is either fixed or evaluated using the error model for the mean mass of the stack (see Appendix \ref{app:error_model_leisng_mass}). This is useful if we want to match the lensing mass error bars that are obtained in a given analysis by just setting them manually.

The simulated cluster count may become zero in one or more richness-redshift bins for a given simulation. In that case, it is not possible to compute the appropriate mean cluster mass for the corresponding bins. As an output of \capish and as an input of the neural density estimators (see the next section), we use the "mass" summary statistics 
\begin{equation}
    \widehat{NM}_{kl} =
\begin{cases}
\widehat{N}_{kl}\times \widehat{M}_{kl} & \text{if } \widehat{N}_{kl} \neq 0 \\
0  & \text{if } \widehat{N}_{kl} = 0,
\end{cases}
\label{eq:NM_kl}
\end{equation}
where $\widehat{N}_{kl}$ is given in \eqreff{eq:count}, and $\widehat{M}_{kl}$ is given in \eqreff{eq:stacked_mass_gamma}. So, \eqreff{eq:NM_kl} is set to 0 when the bin is unfilled, instead of being not defined. This enables us to smooth the summary statistics for extreme cases, which will facilitate the training of neural density estimators. 
\end{itemize}

\begin{table}
\begin{center}
\caption{Fiducial values used in this work.}
\resizebox{0.48\textwidth}{!}{
\begin{tabular}{ccc} 
 Parameters & Default values in \capish & Equations\\
\hline
$(\Omega_m\,, \Omega_b)$ & (0.319, 0.048)&\eqreff{eq:dn_dm_dz}\eqref{eq:sigma2_z1z2_fullsky}\\
$(\sigma_8,\, n_s)$ & (0.813, 0.96)&-\\
$h$ & 0.7&-\\
$(w_0, w_a)$ &(-1, 0)&-\\
\hline
$\Omega_S$ &$\pi/2\, (f_{\rm sky}=1/8)$&-\\
$(\mu^{\lambda}_0\,, \mu^{\lambda}_m\,, \mu^{\lambda}_z)$  & (3.5, 1.72, 0.0)&\eqreff{eq:powerlaw_richness} \\
$(\log_{10}m_0\,, z_0)$ &(14.5, 0.5)& -\\
$\sigma_{\ln \lambda, \rm int}$ & 0.2 & \eqreff{eq:sigma2_lnlambda}\\ 
$\rho$ & 0& \eqreff{eq:covariance_plambda} \\ 
    $(\mu^{\rm WL}_0\,,\mu^{\rm WL}_m\,,\mu^{\rm WL}_z)$ & (0, 1, 0)& \eqreff{eq:log10_mwl_mean}\\ 
   $(\sigma_{\rm WLgal},\,\sigma_{\rm WLint}) $ &  (0.2 or theory-based, 0)&\eqreff{eq:sigma_WL}\\
   $(\sigma_{\rm PZ,0},\,b(z_{\rm true}))$ & (0, 0) &\eqreff{eq:photometric_redshift}\\
$(n_{\rm pur},\, \ln\lambda_p)$ & $(2.5, 1.5)$  &\eqreff{eq:purity}\\
\hline
$(\alpha_{\rm gal}, \beta_{\rm gal}, z_{\rm gal}^0)$ & $(2.0, 1.5,0.5)$  & \eqreff{eq:changnz}\\
$\Gamma$ & 0.7&\eqreff{eq:stacked_mass_gamma}\\
\end{tabular}}
\tablefoot{The halo mass function is taken from \citet{Despali_2015} and the halo bias from \citet{Tinker_2010}. The first block corresponds to the fiducial cosmological parameters. The second block is for generating the cluster catalog from the underlying halo catalog. The last block is for computing summary statistics from the cluster catalog. The parameters $\mu_0^\lambda$, $\mu_m^\lambda$, and $\mu^\lambda_z$ are chosen to mimic the DES Y1 best fit of the cluster mass-richness relation \citep{Abbott2020DESCL}, the latter following a different parameterization.}
\end{center}
\label{tab:default_capish}
\end{table}

\section{Validation}
\label{sec:capish_validation}
\subsection{Validating \capish outputs}
Before any cosmological analysis, we can explore how \capish outputs compare to the theoretical prediction of cluster counts and cluster masses (see Appendix \ref{app:likelihood_standard}). For the redshift bin $0.2 < z < 0.5$ and in seven richness bins, we show in Fig. \ref{fig:capish_outputs_mass} (left panel, upper plot) the bias between the mean \capish-counts (averaged over 100 simulations) and an analytical count prediction in \eqreff{eq:nth_richness_redshift_bins}, which is computed without accounting for selection function with fiducial parameters listed in \tabreff{tab:default_capish} (the underlying halo mass definition used throughout this work is $M_{\rm 200m}$, and we used the \citet{Tinker_2010} halo mass function). For the simplest case (no selection function) \capish counts are in good agreement with the theoretical prediction. Accounting for the selection function\footnote{The fiducial selection function parameters in \tabreff{tab:default_capish} are chosen such that the cluster catalog is increasingly purer at higher richnesses, with $50\%$ purity at richness $\lambda_{\rm obs}=5$. }, we see that adding purity to \capish simulations affects the recovered counts at low richness, by adding fake clusters. We also see that adding photometric redshift (with $\sigma_{\rm PZ}(z) = 0.02(1+z)$, and $b(z) = 0$ in \eqreff{eq:photometric_redshift}) induces a 2-3$\%$ bias with respect to the true redshift case. In Figure \ref{fig:capish_outputs_mass} (left panel, lower plot), we see that the variance of \capish counts is in good agreement with the analytical prediction of the cluster count variance, accounting for Poisson noise and SSC in \eqreff{eq:covariance}.

Figure \ref{fig:capish_outputs_mass} (right panel, upper plot) shows the bias between the mean simulated \capish masses and a mean-mass prediction (computed using \eqreff{eq:mth_richness_redshift_bins}), again with the simplest assumptions (i) no selection function (ii) no scatter for the lensing mass and (iii) $\Gamma = 1$. When \capish is run under these simplest assumptions, the bias is zero. Introducing a slope $\Gamma = 0.7$ produces a $5\%$ bias, and adding noise -- from galaxy shape and/or shot noise and intrinsic scatter (0.2 and 0.05, respectively) -- to the lensing mass (assuming the $\log_{10}m_{\rm WL}$ parameterization) increases this bias to $10\%$. Furthermore, varying the constant correlation parameter $\rho \in \{-0.1, +0.1\}$ causes the mean lensing masses to be biased low and high, respectively. More realistically, \capish allows the use of a mass-dependent function $\rho(m) = 0.3\times \exp{-2(\log_{10}m - 13.3)}$, decreasing with mass, whose effect on mean mass is more pronounced at lower richness (or lower mass) values. This is more realistic with regard to the mass-dependence of $\rho$ \citep{Sunayama2020projection,Wu2022selection,Zhou2024selectionbias}. 

The right panel of Fig. \ref{fig:capish_outputs_mass} (lower plot) shows the dispersion of the \capish mean masses. In the absence of lensing-mass noise, the dispersion is at its lowest level, originating from Poisson and SSC sampling of the halo mass function. When lensing-mass noise is included, the dispersion increases accordingly. We also see that using the error model described in Appendix \ref{app:error_model_leisng_mass} (with $\bar{n}_{\rm gal}=25$ arcmin$^{-2}$, and considering the fitting radius between 1 Mpc and 5 Mpc) increases the errors at low richness and decreases them at high richness, compared to adopting a fixed value for $\sigma_{\rm WL}$. All of these preliminary tests of the \capish simulator ensure that the summary statistics behave as expected, as for their estimators and their covariances. 

\subsection{Probability coverage}
\label{sec:bayesian_robustness}
The posterior prediction is built on top of the \texttt{sbi}\footnote{\url{https://sbi-dev.github.io/sbi/latest/}} Python package \citep{tejerocantero2020sbi}, a flexible toolkit for SBI. The required inputs for \texttt{sbi} Python package are the summary statistics obtained from the \capish simulator. Then, \texttt{sbi} trains neural networks, typically normalizing flows, to learn the posterior $\mathbb P(\widehat{\theta}_{\mathrm{true},k} \mid \widehat{D}_k)$ using $N_{\rm sim}$ simulated pairs $\{\widehat{\theta}_{\mathrm{true},k},\widehat{D}_k\}$ for $k=1, 2, ..., N_{\rm sim}$, where $\widehat{D}_k \sim \mathrm{Simulator}(\widehat{\theta}_{\mathrm{true},k})$.

We assess the performance of the posterior generator across the free parameter space
$\theta = \{\Omega_m, \sigma_8, \mu_0^\lambda, \mu_m^\lambda, \mu_z^\lambda, \sigma_{\ln\lambda, \rm int}\}$.
For the \capish simulator, we consider: (i) a sky area of $\Omega_S = \pi/2$ ($f_{\rm sky}=1/8$); (ii) the \citet{Tinker_2010} halo mass function with the $200m$ mass definition; (iii) pure cluster samples; (iv) the scaling functional form of the scaling relation in \eqreff{eq:powerlaw_richness}; (v) no photometric redshifts; and (vi) we consider the stacked approach for the scatter of weak lensing mass, as explained in Sect. \ref{sec:summary_stats}, and applied to the lensing mass through \eqreff{eq:sigma_wl_stack} with a theory model for $\sigma_{\rm WL}$ (with $\bar{n}_{\rm gal}=25$ arcmin$^{-2}$; see Appendix \ref{app:error_model_leisng_mass}); (vii) using the alternative mass summary statistics in \eqreff{eq:NM_kl}. Summary statistics (counts and modified lensing masses) are computed using the richness bin edges
$\lambda_{\rm obs} = \{20, 30, 40, 60, 100, 200\}$ and redshift bin edges
$z = \{0.2, 0.35, 0.5, 0.7, 1.0\}$. We use the lower richness cut (respectively, redshift) $\lambda_{\rm obs} > 20$ (respectively, $z > 0.2$), to be consistent with DES-Y1 analyzes \citet{McClintock2019masscalibration,Abbott2020DESCL}; the $\lambda_{\rm obs} > 20$ cut is generally used so that it ensures a high-purity cluster sample \citep{Costanzi2019SDSSCL}. Restricting to $z > 0.2$ mimics the conservative cut used in DES cluster-based analyzes \citep{Abbott2020DESCL,Abbott2025DESCLY3}, to prevent the degradation of \texttt{redMaPPer} performance at low redshifts, where the red-sequence galaxy population becomes harder to isolate due to the lack of $u$-band data in the DES analyzes\footnote{The \texttt{redMaPPer} detection in the different DES cluster analyzes used the $g$, $r$, $i$, and $z$ bands. For $z < 0.2$ galaxies, the distinctive break features at $\sim 4,000\, \r{A}$ fall in the $u$-band.}. We note that robust detection of low-redshift clusters below $z = 0.2$ will be feasible with the Rubin LSST, since $u$-band imaging will be available over $18,000$ square degrees of the Southern sky. For the fiducial parameters, the total number of clusters within the considered richness and redshift bins is $\sim 77,000$. For the first redshift bin, $N_{z_1,\lambda} \sim \{6200, 1900, 1100, 340, 45\}$. For the second redshift bin, $N_{z_2,\lambda} \sim \{10^4, 2900, 1500, 400, 45\}$. For the third redshift bin, $N_{z_3,\lambda} \sim \{1.6\times 10^4, 4200, 1900, 430, 34\}$ and for the fourth redshift bin $N_{z_4,\lambda} \sim \{2.3\times 10^4, 4900, 1900, 320, 17\}$. The typical binned $\langle \log_{10}(M_{\rm 200m}/M_\odot) \rangle$ is rather stable with redshift and goes from 14.2 (low richness bin) to 15 (high richness bin). Our upper richness cut, $\lambda_{\rm obs} < 200$, ensures that all massive simulated clusters are included in the counts\footnote{This holds for simulations run with the fiducial parameters in \tabreff{tab:default_capish}.}. While our analysis focuses on the sensitivity of cluster abundance to growth-of-structure parameters (such as $\Omega_m$ and $\sigma_8$) within the standard Lambda cold dark matter ($\Lambda$CDM) paradigm, galaxy clusters also provide a powerful probe of extensions beyond $\Lambda$CDM. These include scenarios with massive neutrinos \citep{Bohringer2016neutrinos}, modified gravity \citep{Cataneo2015fRclusters}, and primordial nonGaussianity \citep{Robinson2000loclPngclusters}. Testing such extensions typically requires probing the abundance of rarer systems, i.e., very massive and/or high-redshift clusters. Since the last richness bin remains relatively broad (spanning $\lambda_{\rm obs} = 100$ to $\lambda_{\rm obs} = 200$), it could be refined in future work to better capture the high-mass tail of the halo mass function. This would allow us to more fully exploit SBI for constraining extended cosmological models.

We generate $N_{\rm sim} = 60{,}000$ simulations that cover the parameter priors
$\Omega_m \sim \mathcal{U}(0.2, 0.45)$,
$\sigma_8 \sim \mathcal{U}(0.6, 0.95)$,
$\mu_0^\lambda \sim \mathcal{U}(3, 4)$,
$\mu_m^\lambda \sim \mathcal{U}(1.3, 2.1)$,
$\mu_z^\lambda \sim \mathcal{U}(-0.7, 0.7)$, and
$\sigma_{\ln\lambda, \rm int} \sim \mathcal{U}(0.1, 0.5)$, which encompass the fiducial values listed in \tabreff{tab:default_capish}. This stage took approximately two hours of CPU time.

We train the posterior generator using the neural posterior estimator (\texttt{NPE}) method, using alternative cluster counts in the dedicated richness-redshift bins (this configuration is labeled $\texttt{count}$; see \eqreff{eq:count}), mean lensing masses multiplied by the cluster counts (labeled $\texttt{Nm}$; see \eqreff{eq:NM_kl}), and the combination of the two (labeled $\texttt{count$\_$Nm}$). The SBI training for each setup took approximately 25 minutes of CPU time. After training, the \texttt{sbi} package returns a prediction for the posterior distribution for the six parameters $\theta = \{\Omega_m, \sigma_8, \mu_0^\lambda, \mu_m^\lambda, \mu_z^\lambda, \sigma_{\ln\lambda, \rm int}\}$, given an observed data vector $\widehat{D}$ (being cluster counts, cluster masses, or a combination of the two), For a given data vector $\widehat{D}$, approximately 500,000 samples of the posterior $\widehat{\theta} \sim\mathbb P(\cdot\mid \widehat{D})$
can be drawn within a second.

Validating the accuracy of the posterior generator is critical. To this end, we assess its performance using the coverage calibration test, a key diagnostic of the posterior that tests if the reported posterior credible sets have their intended probabilistic meaning.

Given a dataset $\widehat{D}$ simulated at a parameter point $\widehat{\theta}_{\mathrm{true}}$ drawn from the prior (i) we compute the corresponding posterior estimate $\mathbb P(\widehat{\theta} \mid \widehat{D})$ from the trained NPE, and (ii) we extract a $\gamma$-credible set\footnote{A $\gamma$-credible set is a set that contains the parameter value with probability $\gamma$, e.g., the interval between the $\gamma/2$-quantile and the $1-\gamma/2$-quantile in 1D.} $\mathrm{Cred}_\gamma(\widehat{D})$ for this estimated posterior. For a nominal coverage probability $\gamma$, the Bayesian coverage expresses as
\begin{align}
p_\gamma(\widehat{D}) &:= \mathbb P\!\left(\widehat{\theta}_{\mathrm{true}} \in \mathrm{Cred}_\gamma(\widehat{D}) \mid \widehat{D}\right)\\
&\ = \mathbb E_{\widehat{\theta}_{\rm true}}[\mathds{1} \{\widehat{\theta}_{\mathrm{true}} \in \mathrm{Cred}_\gamma(\widehat{D})\} \mid \widehat{D}].
\label{eq:probability_coverage_Pgamma}
\end{align}
A posterior estimate is calibrated in the Bayesian sense if $p_\gamma(\widehat{D})=\gamma$ for any $\widehat{D}$, that is, the true parameter value $\widehat{\theta}_{\mathrm{true}}$ should fall within the estimated posterior $\gamma$-credible set a fraction $\gamma$ of the time, and so for any observed data $\widehat{D}$. Under mild assumptions, this calibration is equivalent to the estimated posterior being equal to the true posterior.
However, we do not have access to the true posterior by construction, and therefore the computation of $p_\gamma$ from simulated pairs would require some binning of the observations. The common method is to average $p_\gamma(\widehat{D})$ also on the data, i.e., over the simulated pairs $\{\widehat{\theta}_{\mathrm{true}, k}, \widehat{D}_k\}$, giving
\begin{equation}
p_\gamma := \mathbb P\!\left(\widehat{\theta}_{\mathrm{true}} \in \mathrm{Cred}_\gamma(\widehat{D})\right) = \mathbb E_{\widehat{\theta}_{\rm true}, \widehat{D}}[\mathds{1} \{\widehat{\theta}_{\mathrm{true}} \in \mathrm{Cred}_\gamma(\widehat{D})\}],
\label{eq:probability_coverage_Pgamma}
\end{equation} 
called expected coverage probability (ECP). We say that the posterior is calibrated with respect to ECP when $p_\gamma=\gamma$. For information, a sub-identity $p_\gamma < \gamma$ (respectively, super-identity $p_\gamma > \gamma$) calibration curve reveals an underestimation (respectively, overestimation) of the uncertainty.

This calibration test allows us to detect some potential failures in the posterior estimation. For example, due to the continuous nature of the normalizing flows employed in the neural density estimation (NDE), the posterior estimate can be highly inaccurate near sharp posterior features such as prior edges (for example, \citealt{Reza2024sbi}). This leads to an underestimation of the probability density near the prior boundaries such that a probability coverage test would exhibit a mild tendency toward overconfidence. First, we do not account for this effect and compute the ECP over the full prior range, for the three configurations (counts, mean masses, and their combination), as shown in Fig. \ref{fig:prob_coverage}, where it appears that $p_\gamma <\gamma$, i.e., the uncertainties are indeed underestimated. We therefore recompute the same probability coverage for the $\texttt{count\_Nm}$ configuration (combination of counts and mean masses, the most relevant), but restricting the evaluation to fewer points $\widehat{\theta}_{\rm true} \in [\theta_{\min} +\Delta \theta/2, \theta_{\max} - \Delta \theta/2]$, where $\theta_{\min}$ (respectively,  $\theta_{\max}$) is the lower (respectively, upper) bound of the parameter prior, $\Delta \theta$ represents $\sim$ 5$\%$ of the prior size, and we ``mask" posterior samples accordingly by removing samples outside this updated prior range. 
As shown in the left panel of Fig. \ref{fig:coverage_plot_restricted} (for the coverage plot, we consider 1000 pairs $\{\widehat{\theta}_{\mathrm{true}, k}, \widehat{D}_k\}$ with $\widehat{\theta}_{\mathrm{true}, k}$ within the restricted prior region, and posteriors are estimated with 50,000 samples), the resulting calibration curve is closer to identity, emphasizing that this prior problem can be solved by (i) training on a broader prior, (ii) restricting the posterior to a tighter prior range, chosen to match the analysis requirement.

Additionally, to test not only the marginalized but also the joint posterior, we compute an alternative ECP quantity by using the TARP method (Tests of Accuracy with Random Points; \citealt{Lemos2023sbi}), shown in the left panel of Fig. \ref{fig:coverage_plot_restricted}, also close to identity, ensuring good calibration with respect to this metric. Although they do not guarantee the full validity of the posterior estimation, these tests are effective in detecting potential miscalibration, and all pass successfully here.

\subsection{Cosmological analysis: Internal validation}
\label{sec:intern_valid}
Given the configuration $\texttt{count\_Nm}$, and considering 500 pairs $\{\widehat{\theta}_{\mathrm{true}, k},\widehat{D}_k\}$, we compute the estimated posterior mean $\mathrm{E}[\widehat{\theta}_{k} \mid \widehat{D}_k]$ that we compare to the true values $\widehat{\theta}_{\mathrm{true}, k}$. The comparison is shown in Fig. \ref{fig:parameter_bias}. This consistency test shows that the recovered cosmological parameters are efficiently recovered from the posterior, with some scatter around the true value, with no strong apparent bias.  

\begin{figure*}
\begin{center}
\includegraphics[width=0.49\textwidth]{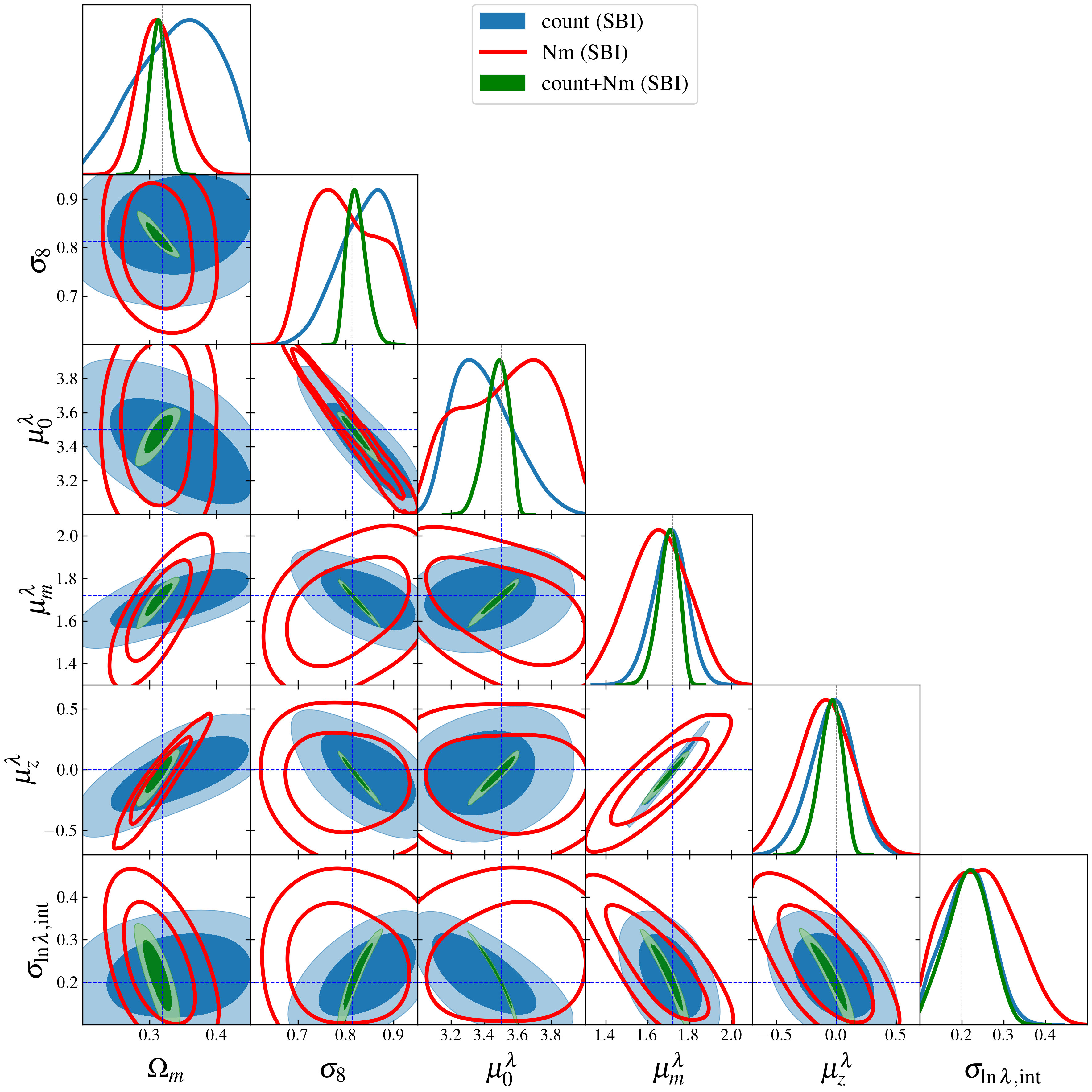}
\includegraphics[width=0.49\textwidth]{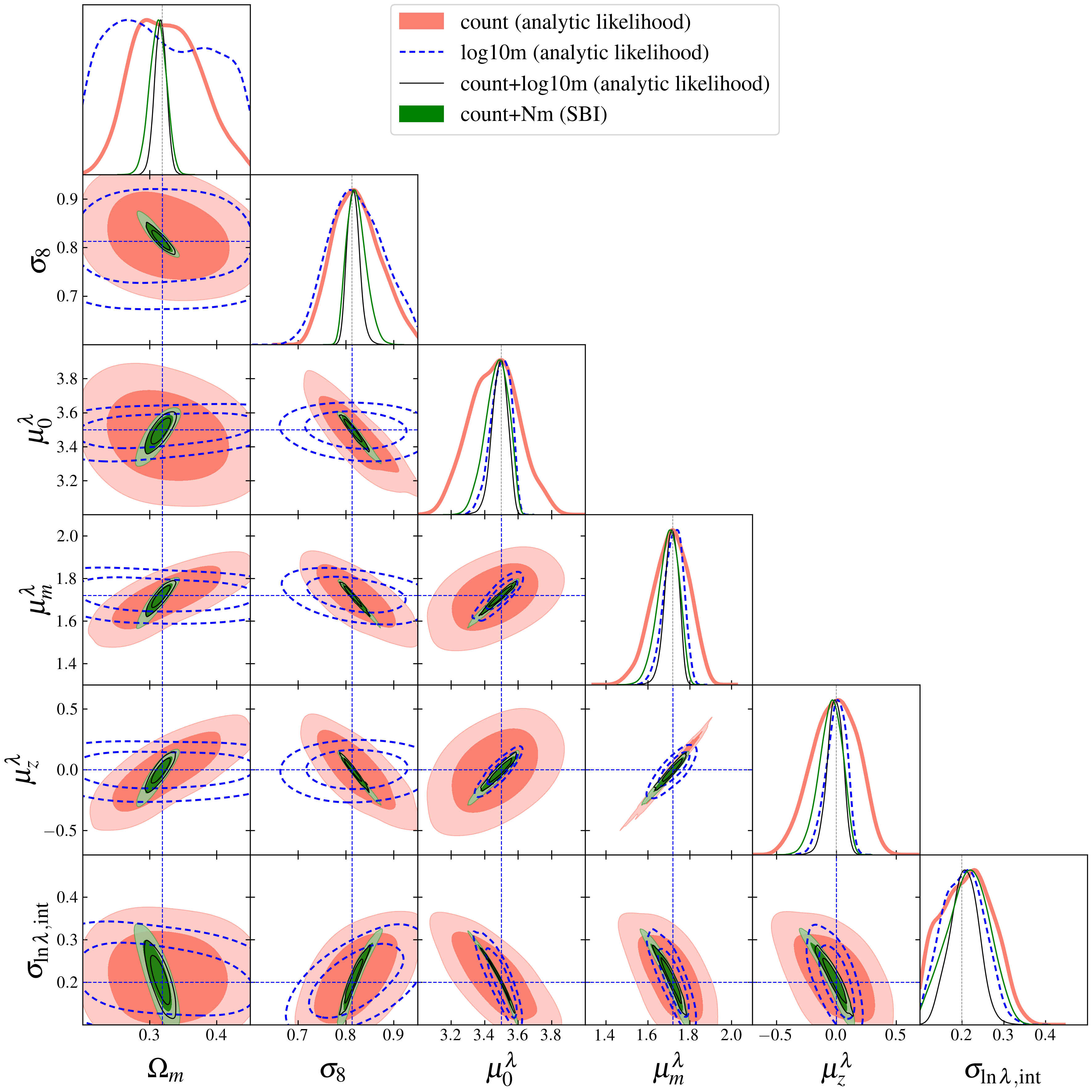}
\caption{\textit{Left}: Posterior distributions obtained from the trained posterior generator for the three configurations \texttt{count}, \texttt{Nm}, and \texttt{count\_Nm}. \textit{Right}: Corresponding posteriors from the analytic likelihood approach, sampled with MCMC for \texttt{count}, \texttt{log10m}, and \texttt{count\_log10m} (the SBI \texttt{count\_Nm} results are overplotted for comparison). }
\label{fig:posterior_mcmc_sbi}
\end{center}
\end{figure*}

We study now the shape of the posterior, for which we generate 500 simulations at the same fiducial parameter values (in \tabreff{tab:default_capish}).
By averaging the data vector over these realizations, we obtain an effectively noiseless data vector, which we use to perform consistency tests of the Bayesian inference.
Figure \ref{fig:posterior_mcmc_sbi} shows the resulting posterior distributions\footnote{Posteriors are displayed with \texttt{GetDist} \citep{Lewis2025getdist}.} (using \texttt{count}, \texttt{Nm}, or \texttt{count\_Nm}) for the six parameters (the best fits are reported in \tabreff{tab:best_fits}).
In all cases, we find good agreement between the inferred posteriors and the fiducial parameter values, the latter lying within the $1\sigma$ region of each posterior distribution, as expected from the test in Fig. \ref{fig:parameter_bias}. 

As mentioned before, it is possible that cluster number counts of \capish simulations display one or more empty bins, making the computation of the mean mass impossible (that is why we choose the alternative lensing mass statistics). In Appendix~\ref{app:emptybins} we explore the effect of removing the simulations that contain at least one empty cluster-count bin in the training. In this case, $\log_{10}M_{kl}$ can be computed and then used in the NDE training. In Fig. \ref{fig:empty_bin_map} we show the fraction of nonremoved objects after masking over empty count bins, displaying an inhomogeneous shape over the prior space (with $\sim$40$\%$ remaining simulations in the bottom left corner of the prior region). This masking induces a complex, implicit prior on the posterior and breaks the desired transparency of the pipeline to empty bins, and shifts the posterior (as shown in Fig. \ref{fig:mcmc_masked}). This confirms that the alternative summary statistics $\texttt{Nm}$ in \eqreff{eq:NM_kl} are particularly reliable for this task.

\begin{figure*}
    \centering
    \includegraphics[width=0.49\linewidth]{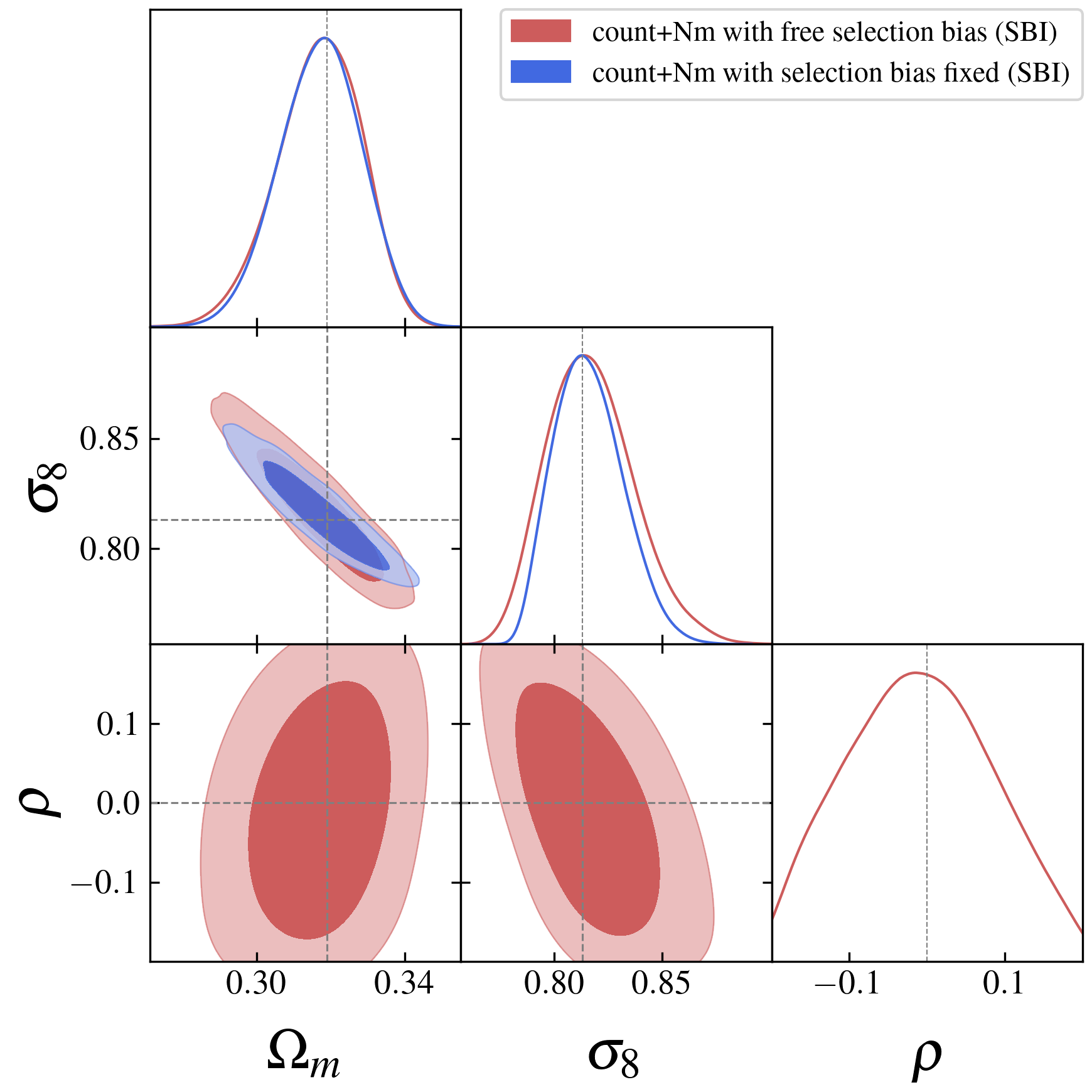}
    \includegraphics[width=0.49\linewidth]{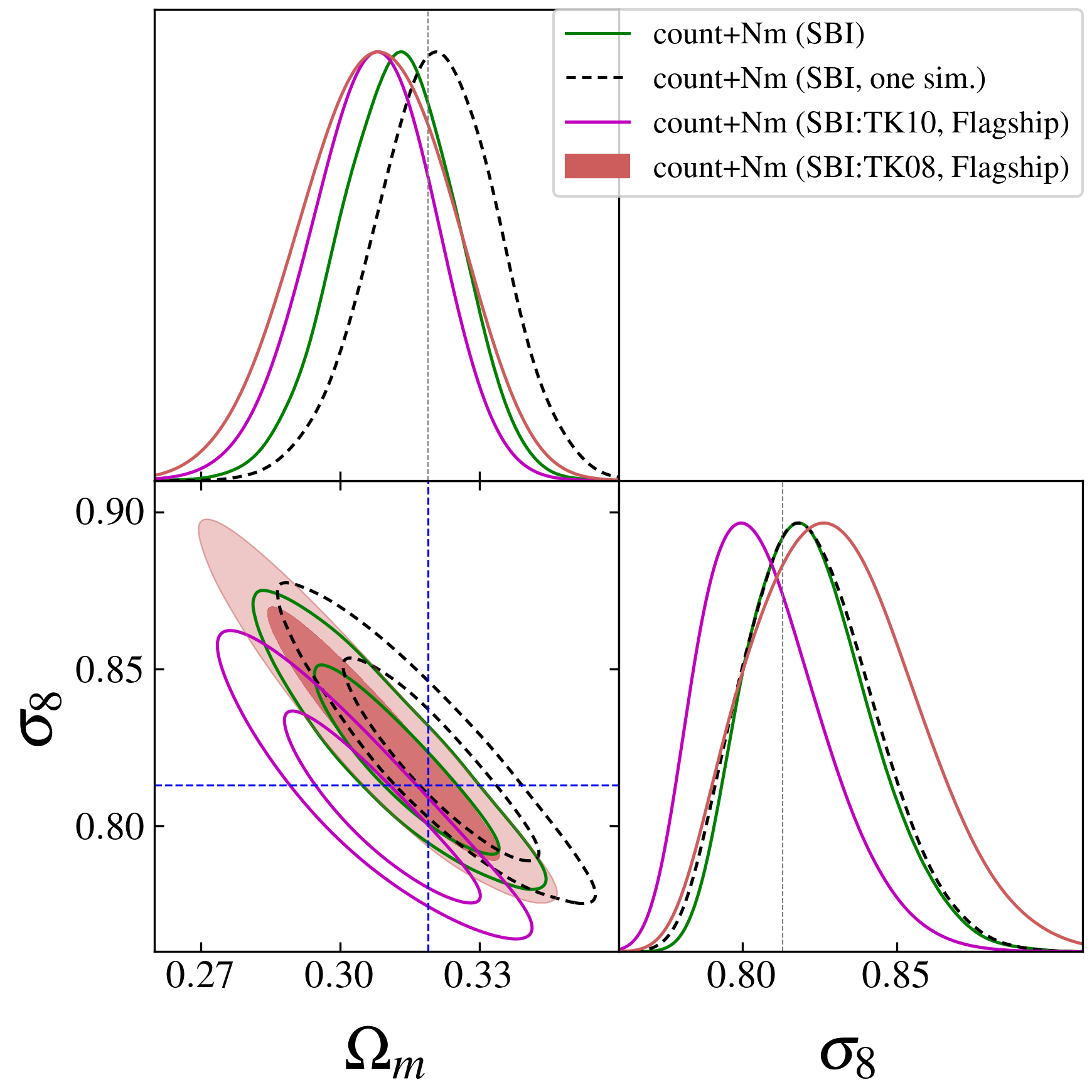}
    \caption{\textit{Left}: Posterior distributions obtained on a noiseless data vector with the \texttt{count\_Nm} setup; the blue posterior is obtained by training on only 6 parameters, and the red is trained adding the lensing mass-richness correlation parameter $\rho$. \textit{Right}: Posterior distribution over the $\Omega_m$ and $\sigma_8$ parameters.
(i) \texttt{count\_Nm} applied to the noiseless data vector, 
(ii) \texttt{count\_Nm} applied to a single fiducial simulation, 
(iii) \texttt{count\_Nm} applied to the Flagship simulation with richness drawn from \capish, trained on \citet{Tinker_2010} halo mass function and (iv) same as (iii), but trained on \citet{Tinker2008hmf} halo mass function.}
    \label{fig:mcmc_flagship}
\end{figure*}

We test how the selection-bias parameter $\rho$ in \eqreff{eq:covariance_plambda} -- which quantifies the correlation between observed weak-lensing masses and optical richness -- impacts cosmological parameter constraints. To this end, we adopt a different noise model than in previous analyzes: instead of applying lensing-mass scatter to the stacked mass, the scatter is applied to individual lensing mass measurements.
As a baseline, we first train a new \texttt{count\_Nm} NDE with $\rho$ being fixed. We then train a second \texttt{count\_Nm} NDE that includes $\rho \sim \mathcal{U}(-0.2, 0.2)$. The resulting posterior samples, applied to the noiseless data vector (computed with $\rho=0$), are shown in Fig. \ref{fig:mcmc_flagship} (left panel), marginalized over the parameters $\mu_0^\lambda$, $\mu_m^\lambda$, $\mu_z^\lambda$, and $\sigma_{\ln \lambda, {\rm int}}$.
Allowing $\rho$ to vary leads to slightly broader constraints on $\sigma_8$, revealing a negative correlation between the selection-bias parameter and $\sigma_8$, and keeping the constraints on $\Omega_m$ unchanged. This example illustrates that \capish incorporates the effect of lensing mass-richness correlation, and that, for the considered prior and current parametrization for $\rho$, including this mass-dependent correlation does not modify significantly the inference of cosmological parameters.

\subsection{Cosmological analysis: Comparison to the explicit-likelihood approach} 
In this subsection we show how the SBI posteriors compare to the posteriors obtained from an explicit-likelihood approach. The elements for the explicit-likelihood approach are detailed in Appendix \ref{app:likelihood_standard}; we define a Gaussian cluster abundance likelihood $\mathcal{L}_{\rm BLC}^{\rm Gauss-SN+SSC}$ in \eqreff{eq:binned_gaussian_likelihood_count} accounting for the Poisson noise \citep{Poisson1837} and SSC \citep{2003_SV_HU}. For the mean lensing mass likelihood $\mathcal{L}_{M_{\rm WL}}$, we adopt a Gaussian model in \eqreff{eq:binned_gaussian_likelihood_mass} for the logarithm of stacked cluster masses, where the covariance is taken to be diagonal. Its diagonal components  (i.e., the dispersion of each stacked mass) are computed assuming a Fisher-like approach on an NFW \citep{Navarro1997nfw} profile model, accounting for the shape noise and shot noise of source galaxies, and the total number of clusters within the stack. 

We draw samples from the parameter posterior distribution with the \texttt{emcee} package \citep{Foreman_Mackey_2013emcee} given by the Bayes theorem $\mathbb P(\theta \mid \mathrm{data}) \propto \mathcal{L}_{\rm tot}(\mathrm{data} \mid \theta)\times\pi(\theta)$, where $\pi(\theta)$ is the prior distribution mentioned earlier. Each Markov chain Monte Carlo (MCMC) run took about 4 hours of CPU time. The results are shown in the right panel of Fig. \ref{fig:posterior_mcmc_sbi}; the explicit-likelihood constraints are consistent with the fiducial parameters, showcasing that the simulation-based noiseless \capish data vectors are not biased between the two methods. The combined constraints and count-only show a comparable shape to the SBI results (left panel of Fig. \ref{fig:posterior_mcmc_sbi}), whereas the explicit-likelihood \texttt{log10m} constraints are very different from the \texttt{Nm} approach, since not the same information is encoded between the two statistics. When combined with counts, however, the discrepancy between SBI and the explicit-likelihood approaches decreases, since the combined constraints show roughly similar posterior widths and correlations as those inferred by SBI. However, we see that explicit-likelihood error bars are slightly smaller, which we expect is from a series of effects; in the explicit-likelihood approach, we adopt a fixed covariance matrix\footnote {Computed at the fiducial cosmology and scaling relation parameters.} for the lensing masses and cluster counts (to speed up the calculations), whereas in the SBI approach, the covariance can vary across simulations. Our implementation of the explicit-likelihood approach neglects correlation between probes, which will tighten parameter posterior and \capish accounts for a sampling dispersion (the intrinsic variation of the mean mass in each richness-redshift bin due to the variation of the number of clusters), which is absent from the explicit-likelihood approach\footnote{For which the error bars are computed as $\sigma_{\rm WLgal}/\sqrt{N}$, where $N$ is the cluster count at fiducial values.}. Moreover, the explicit-likelihood approach assumes variables are Gaussian, whereas \capish will account for any nonGaussian noise in our forward model. Given these effects, it is expected that \capish has slightly broader posteriors than the explicit-likelihood approach.

It is important to note that posterior coverage remains a relevant diagnostic for explicit-likelihood approaches (see, e.g., \citealt{Payerne2023}), as it provides a direct test of the accuracy of the recovered posterior, similarly to SBI methods. However, we do not conduct this test in this study, given the high computational cost of evaluating a single explicit-likelihood posterior (4 hours for a single posterior).

\subsection{Cosmological analysis with external simulated datasets}
As a final validation, we apply our trained posterior generator to the \textit{Euclid} Flagship simulation \citep{Euclid2025flagship}\footnote{Publicly available at \url{https://cosmohub.pic.es}}. Flagship is a large-scale mock galaxy catalog designed to support the scientific exploitation of ESA’s \textit{Euclid} mission. The simulation was run using the \texttt{PKDGRAV3} code \citep{Potter2017PKDG} under a flat-$\Lambda$CDM cosmology, with cosmological parameters close to those inferred by \citet{ade2016planck}. A light cone extending to $z = 3$ was generated on the fly, covering one octant of the sky (approximately $5{,}157$ deg$^2$).
Dark matter halos were identified using the \texttt{rockstar} algorithm \citep{Behroozi2013rockstar}. From these halos with true masses $m_{\rm halo}$ and true redshift $z_{\rm halo}$, we construct a cluster catalog by assigning an observed richness $\lambda_{\rm obs}$ to each Flagship halo using \eqreff{eq:powerlaw_richness} and the fiducial parameters listed in \tabreff{tab:default_capish}. Individual lensing masses $m_{\rm WL}$ are taken to be equal to the halo masses $m_{\rm halo}$, with stochasticity applied only at the stacked level through \eqreff{eq:sigma_wl_stack}, when constructing the stacked lensing mass matrices.
Figure \ref{fig:mcmc_flagship} (right panel) shows the posterior distributions for the cosmological parameters $(\Omega_m, \sigma_8)$ obtained with our SBI framework using three different data vectors: (i) a noiseless data vector, (ii) a single \capish simulation, and (iii) the Flagship simulation with \capish-like richnesses. In the case of the single \capish realization, the posterior is slightly shifted, primarily due to statistical noise in the data. For the SBI applied to the Flagship catalog when using the \citet{Tinker_2010} [TK10], we find a good agreement with the fiducial cosmology for the marginalized posteriors, but still with a $\sim 2\sigma$ tension in the $\Omega_m,\sigma_8$ plane.
This residual tension can be attributed to the fact that we performed our training assuming the \citet{Tinker_2010} halo mass function, which is different from the effective mass function measured in the Flagship simulation \citep{Euclid2025flagship}, whose underlying halo mass function was found to be closer to the \citet{Tinker2008hmf} [TK08] implementation (see Appendix \ref{app:flagship_comparison} and the right panel of Fig. \ref{fig:coverage_plot_restricted}). We repeat the training this time using the TK08 halo mass function, which enables us to recover the good cosmological parameters with no significant tension (below $1\sigma$; see the right panel of Fig. \ref{fig:mcmc_flagship}). In practice, the underlying Flagship halo mass function is unknown and thus may differ from TK08 or TK10. Incorporating additional nuisance parameters in the training (denoting any possible deviation from TK08 or TK10) with appropriated priors could help mitigate this discrepancy \citep{Wu2010uncertainties,Cunha2010hmf,Abbott2020DESCL,Artis2021hmfcalibration}.

\section{Conclusions}
\label{sec:conclusions}

In this work, we have presented \capish, a simulation-based inference framework for the cosmological analysis of galaxy cluster datasets. Rather than relying on explicit likelihood modeling, \capish learns posterior distributions directly from forward-modeled cluster observables — specifically cluster number counts and mean weak-lensing masses in richness–redshift bins — naturally capturing nonGaussian features of the data that analytic likelihood approaches cannot. The forward model incorporates Poisson sampling noise and super-sample covariance using log-normal variables, extending the approach of \citet{Payerne2024unbinnedSSC}, as well as lensing mass scatter, selection effects through purity and selection biases, and an alternative lensing mass statistic — given by the product of count and mean lensing mass — to regularize the neural density estimator training. \capish performs posterior inference using neural posterior estimation within the \texttt{sbi} package \citep{tejerocantero2020sbi}.

The code has been validated by comparing its outputs (at fiducial values in \tabreff{tab:default_capish}) against analytical predictions, confirming that the simulated cluster counts accurately reproduce the expected SSC variance. The flexibility of the selection function implementation has been tested, and the impact of lensing mass–richness covariance on recovered masses has been assessed. Bayesian coverage tests indicate that the posteriors are unbiased and provide robust uncertainty estimates for both cosmological and scaling-relation parameters. We have also examined how the effective lensing mass–richness correlation parameter \citep{Wu2022selection,Zhang2024propertycov}, one of the major systematic effects in cosmology with optically selected clusters, propagates into the cosmological posteriors. Comparisons with explicit-likelihood analyzes \citep{Payerne2023} show good overall agreement, with the broader SBI posteriors reflecting the increased realism of the forward model. Another advantage of \capish for simulation-based inference is the computational time, as \capish generates simulations and trains a posterior generator within 2.5 hours (which samples a posterior within a second), whereas one MCMC posterior sampling based on an explicit likelihood takes 4 hours. As a final validation, \capish has been applied to the \textit{Euclid} Flagship dataset \citep{Euclid2025flagship} with \capish-assigned richness, recovering the Flagship cosmological parameters within $1\sigma$. This framework offers a flexible and robust alternative for analyzes of surveys such as DES, \textit{Euclid} \citep{laureijs2011euclid}, and LSST \citep{LSST}, and we propose it as a test bench for validating explicit-likelihood cosmological pipelines and assessing their internal consistency. 
 
\begin{acknowledgements}
    The authors thank the anonymous reviewer for their insightful comments and suggestions. We gratefully acknowledge support from the CNRS/IN2P3 Computing Center (Lyon - France) for providing computing and data-processing resources needed for this work. We thank the developers and maintainers of the following software tools used in this work: \texttt{sbi} \citep{tejerocantero2020sbi}, \texttt{NumPy} \citep{vanderWaltnumpy}, \texttt{SciPy} \citep{jonesscipy}, \texttt{Matplotlib} \citep{Hunter2007matplotlib}, \texttt{GetDist} \citep{Lewis2025getdist}, \texttt{emcee} \citep{Foreman_Mackey_2013emcee}, \texttt{Jupyter} \citep{jupyter}, LSST DESC CCL \citep{Chrasi2019ccl} and LSST DESC CLMM \citep{Aguena2021clmm}.
\end{acknowledgements}

\bibliographystyle{aa}
\bibliography{main.bib} 
\begin{appendix}
\section{SSC validation at the level of the halo mass distribution}
\label{app:ssc_validation}
\begin{figure}
    \centering
\includegraphics[width=0.49\textwidth]{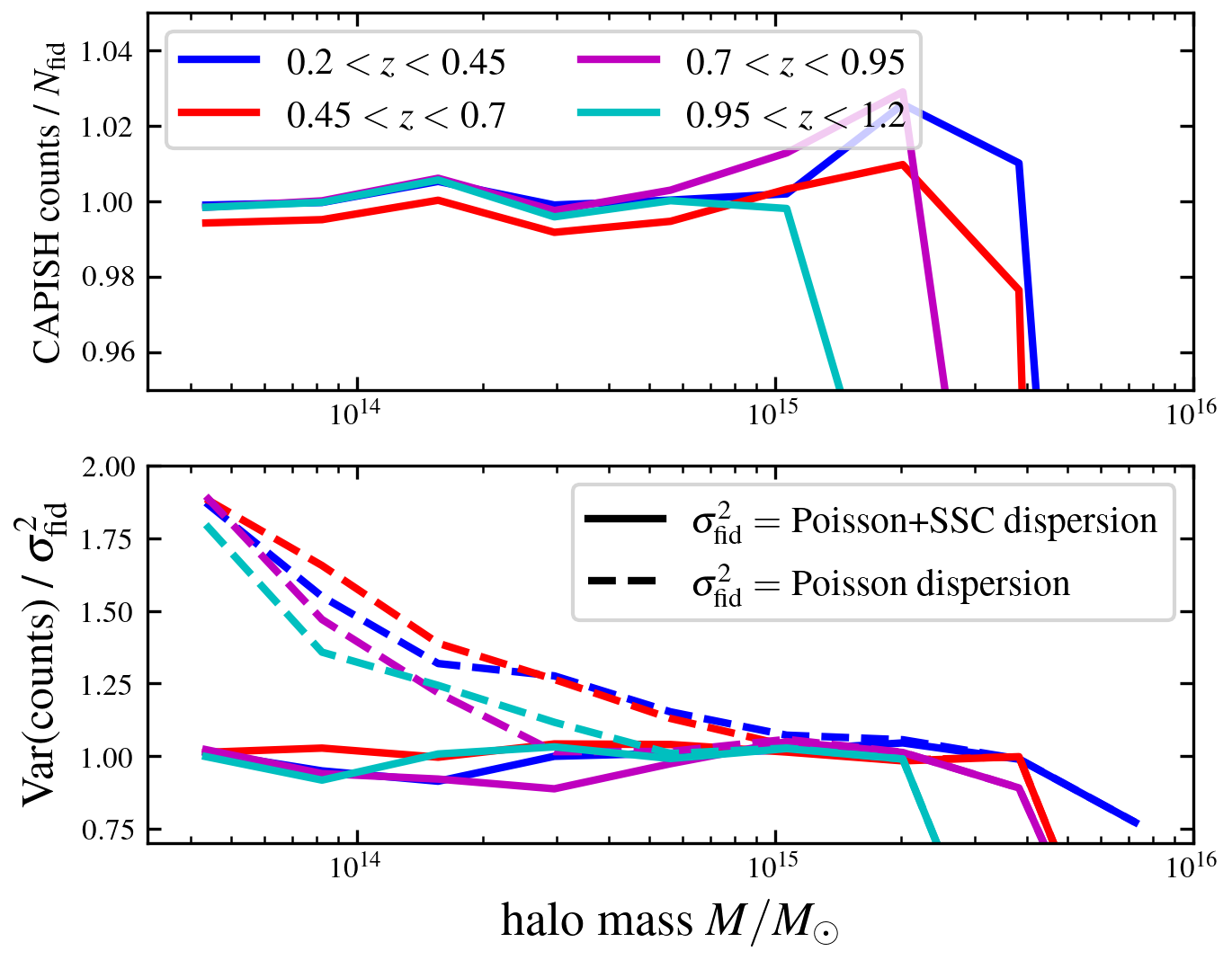}
    \caption{\textit{Top}: Mean \capish halo count compared to analytical prediction. \textit{Bottom}: Variance of \capish halo counts compared to Poisson-only or Poisson+SSC variance analytical prediction.}
    \label{fig:bias_mass_redshift}
\end{figure}

We show in the top panel of Fig. \ref{fig:bias_mass_redshift} the comparison between the mean \capish halo count in bins of true mass and true redshift, and the standard prediction based on \citet{Tinker_2010} halo mass function. The lower panel compares the variance of \capish counts in the same mass and redshift bins, to the Poisson-only variance prediction, and the full Poisson+SSC variance prediction.
\section{Error model for lensing mass}
\label{app:error_model_leisng_mass}
\begin{figure*}
    \centering
\includegraphics[width=0.49\textwidth]{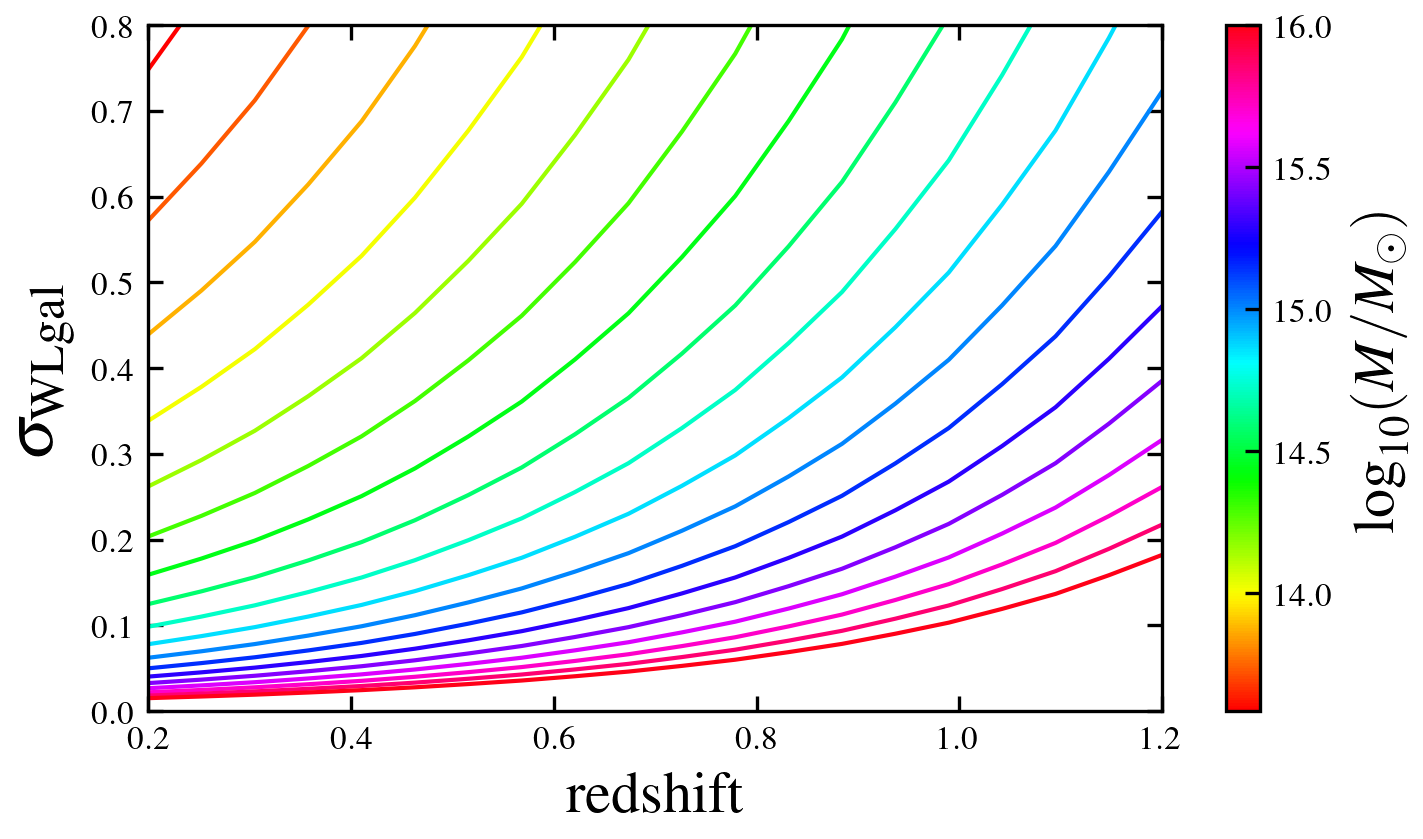}
\includegraphics[width=0.49\textwidth]{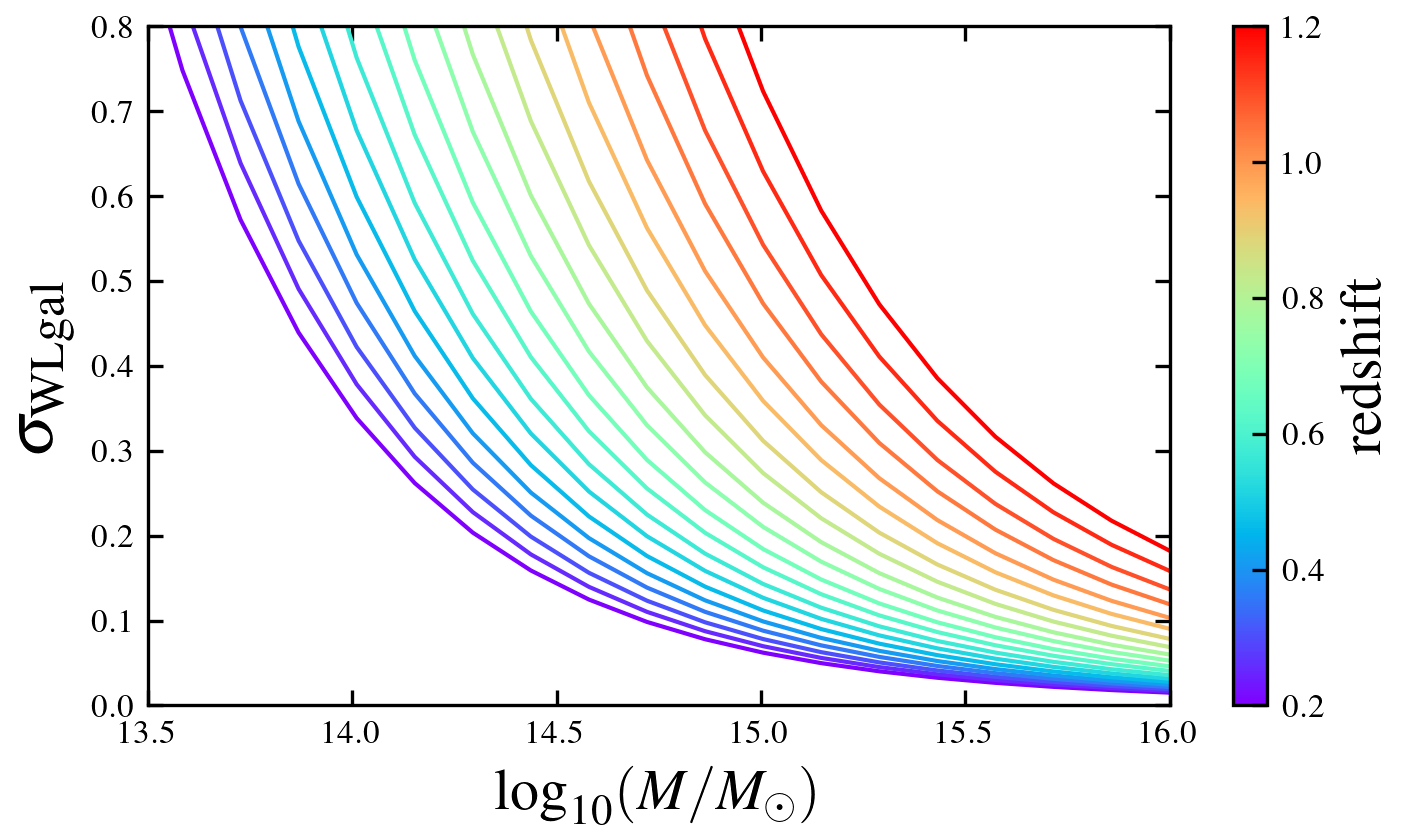}
    \caption{\textit{Left}: Dispersion on the $\log_{10}m_{\rm WL}$ as a function of redshift for different halo mass values. \textit{Right}: Dispersion on the $\log_{10}m_{\rm WL}$ as a function of halo mass for different redshift values.}
    \label{fig:sigmaWL}
\end{figure*}

For a single halo with mass $m$ and redshift $z$, the Fisher information on the weak lensing mass is given by
\begin{equation}
    F_{MM} =
    \int_{R_{\rm min}}^{R_{\rm max}}
    \sigma^{-2}_{\Delta\Sigma}(R)
    \left(\frac{\partial \Delta\Sigma(R)}{\partial M}\right)^2\, dR.\label{eq:fisher_mass_single}
\end{equation}
In the above equation, the integral runs over the available radial range that is used for the weak lensing mass fitting. The quantity $\Delta\Sigma(R)=\Delta\Sigma_{\rm NFW}(R|M,c)$ denotes the predicted excess surface density for an NFW \citep{Navarro1997nfw} halo of halo mass $m$, concentration $c(m,z)$, taken to follow the \citep{Duffy2008cM} concentration mass-relation\footnote{For the prediction of the \citet{Duffy2008cM} concentration-mass relation, we use CCL \citep{Chrasi2019ccl}, and for the prediction of $\Delta\Sigma_{\rm NFW}(R|M,c)$, we use the LSST DESC Cluster weak Lensing Mass Modeling library (CLMM, \citealt{Aguena2021clmm}).}. The per-radius dispersion of the excess surface density profile entering \eqreff{eq:fisher_mass_single} has several sources listed in Sect. \ref{sec:capish_galaxy_cluster_cat}. We focus on the shape and shot noise of the source galaxy sample, given by
\begin{equation}
    \sigma^2_{\Delta\Sigma}(R) =
\frac{\Sigma_{\mathrm{crit}}^2(z_{\rm cl})\, \sigma_\epsilon^2}
         {n^{\rm bgd}_{\mathrm{gal}}(z_{\rm cl})\, 2\pi R},\label{eq:shape_noise_dispersion}
\end{equation}
where $\sigma_\epsilon$ is the intrinsic galaxy shape noise and
$n^{\rm bgd}_{\mathrm{gal}}(z)$ is the surface density of
background galaxies behind a cluster at redshift $z$, given by
\begin{equation}
    n^{\rm bgd}_{\mathrm{gal}}(z_{\rm cl})
    = \bar{n}_{\mathrm{gal}}
    \frac{
    \int_{z_{\rm cl}+0.2}^{\infty} n_{\rm gal}(z_s)\,dz_s
    }{
    \int_{0}^{\infty} n_{\rm gal}(z_s)\,dz_s
    }.
\end{equation}
where the \citet{Chang2013density} normalized redshift distribution is given by
    \begin{equation}
        n_{\rm gal}(z_s) \propto z^{\alpha_{\rm gal}}_s \exp\left(-\left(\frac{z_s}{z^0_{\rm gal}}\right)^{\beta_{\rm gal}}\right),
        \label{eq:changnz}
    \end{equation}
where $\alpha_{\rm gal}$, $\beta_{\rm gal}$ and $z^0_{\rm gal}$ are given in \tabreff{tab:default_capish}. The effective critical surface density is averaged over the source
redshift distribution $n(z_s)$, such as
\begin{equation}
\Sigma_{\mathrm{crit}}^{-2}(z_{\rm cl}) =
    \frac{
    \int_{z_{\rm cl}+0.2}^{\infty} n_{\rm gal}(z_s)\, \Sigma_{\mathrm{crit}}^{-2}(z_{\rm cl}, z_s)\, dz_s
    }{
    \int_{z_{\rm cl}+0.2}^{\infty} n_{\rm gal}(z_s)\, dz_s
    }.
\end{equation}
Then, we get that
\begin{equation}
    \sigma_{\rm WL}(m,z) = \frac{\sqrt{F_{MM}^{-1}}}{\ln(10)m},
\end{equation}
represented in Fig. \ref{fig:sigmaWL} as a function of mass and redshift.

\section{Detected/full halo population after minimal richness cut}
\begin{figure}
    \centering
\includegraphics[width=0.48\textwidth]{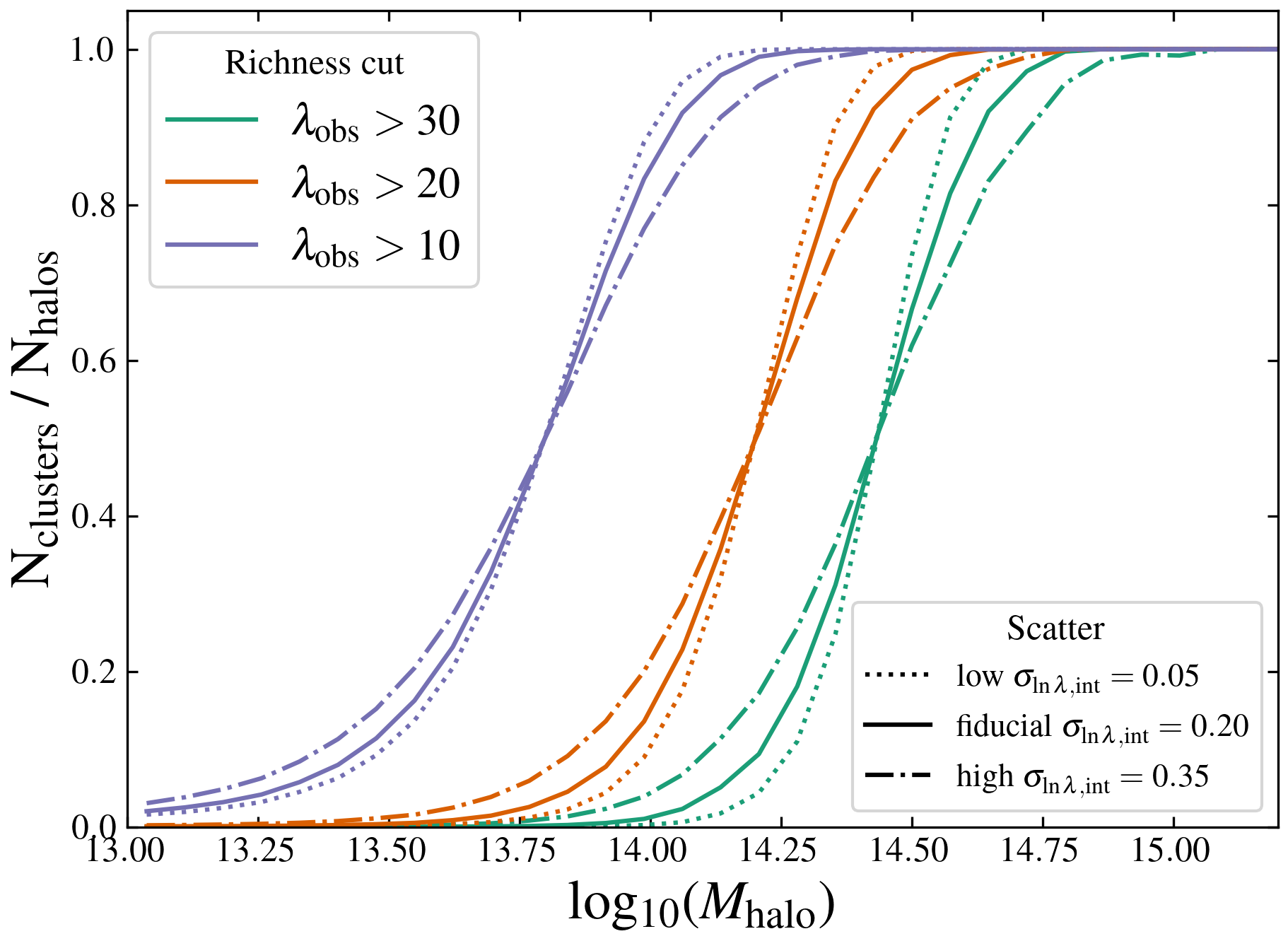}
    \caption{Completeness of the simulated cluster catalog as a function of halo mass for different observed richness thresholds, and intrinsic richness–mass scatter $\sigma_{\ln\lambda, \mathrm{int}}$ in \eqreff{eq:sigma2_lnlambda}. Lower richness thresholds and smaller intrinsic richness–mass scatter yield higher completeness at fixed mass, with all curves asymptoting to unity at high mass.}
    \label{fig:completeness}
\end{figure}

Figure \ref{fig:completeness} illustrates the fraction of detected \capish halos for 3 different minimal cuts in cluster's richness, and for 3 different values of the cluster richness-mass relation intrinsic scatter $\sigma_{\ln \lambda, \rm int}$ in \eqreff{eq:sigma2_lnlambda}.
\section{Alternative interpretation of completeness}
\label{app:alternative_completeness}
In this appendix we discuss an alternative implementation of the survey selection function. In addition to purity, denoting the contamination level of the cluster catalog by spurious detections (implemented in \capish), we discuss here the possibility of including an independent completeness function, representing the fraction of the underlying dark matter halo population that is systematically missing from our dataset \citep{Mantz2010CCmethodunbinned,Mantz2019miss}. Following the formalism in \citet{Aguena2018completenesspurity,Payerne2025cosmodc2,Lesci2022KIDSCL,Lesci2025kidscl}, the expected cluster number count per interval of halo mass, true redshift, observed redshift, and observed richness is given by
\begin{equation}
\begin{split}
    \frac{\partial^4 N(m,z,\lambda_{\rm obs},z_{\rm obs})}{\partial \lambda_{\rm obs} \, \partial m \, \partial z_{\rm obs} \, \partial z}
    = \frac{\partial^2 N(m,z)}{\partial m \, \partial z}
    \, \frac{c(m,z)}{p(\lambda_{\rm obs}, z_{\rm obs})}\,\times\\ P(\lambda_{\rm obs} \mid m,z)
    \, P(z_{\rm obs} \mid z),
    \label{eq:N_c_p}
\end{split}
\end{equation}
where $\frac{\partial^2 N(m,z)}{\partial m \, \partial z}$ denotes for the mass and redshift distribution of the halo population (encoded in the halo mass function and comoving volume), $P(\lambda_{\rm obs} \mid m,z)$ is the cluster scaling relation, $P(z_{\rm obs} \mid z)$ is the observed redshift-true redshift relation, and $p(\lambda_{\rm obs}, z_{\rm obs})$ (respectively, $c(m,z)$) is the purity (respectively, the completeness) of the cluster finder algorithm. From \citet{Aguena2018completenesspurity}, completeness can be modeled as purity with a smoothed step function 
\begin{align}
\label{eq:completeness}
    c(m,z) &= \frac{(m/m_c)^{n_{\rm comp}(z)}}{1+(m/m_c)^{n_{\rm comp}(z)}}.
\end{align}
From a simulation point of view, for each halo with mass and redshift $m, z$, we generate $u\sim U(0,1)$, if $u > c(m, z)$, the halo is removed from the catalog. Along with the implementation of purity in Sect. \ref{sec:capish_galaxy_cluster_cat}, completeness reproduces the \eqreff{eq:N_c_p}.
\section{Dealing with empty bins}
\label{app:emptybins}
\begin{figure}
    \centering
    \includegraphics[width=1\linewidth]{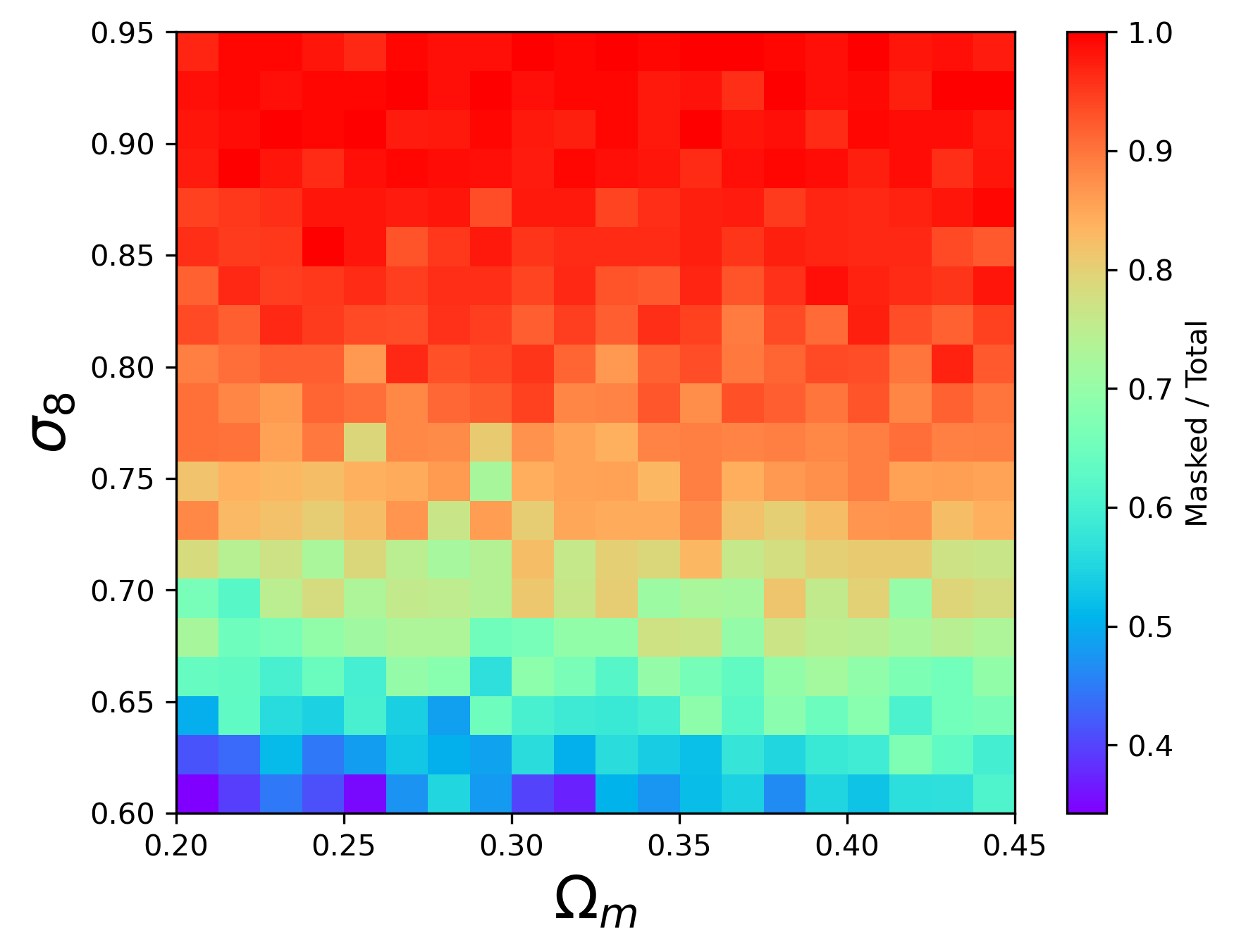}
    \caption{Ratio between the two-dimensional distribution of sampled parameters in the $\Omega_m$–$\sigma_8$ plane and the same distribution after masking simulations that contain at least one empty richness–redshift bin.}
    \label{fig:empty_bin_map}
\end{figure}

In Fig. \ref{fig:empty_bin_map} we show the ratio map between (i) the full set of 30{,}000 parameter samples $\theta_k$ drawn from the priors defined in Sect. \ref{sec:bayesian_robustness} and (ii) the same parameter samples after masking those that produce binned cluster counts with at least one empty bin. We find that most of the excluded samples are concentrated in the lower-left region of the two-dimensional parameter space, corresponding to low values of $\Omega_m$ and $\sigma_8$.

\begin{figure}
    \centering
    \includegraphics[width=0.9\linewidth]{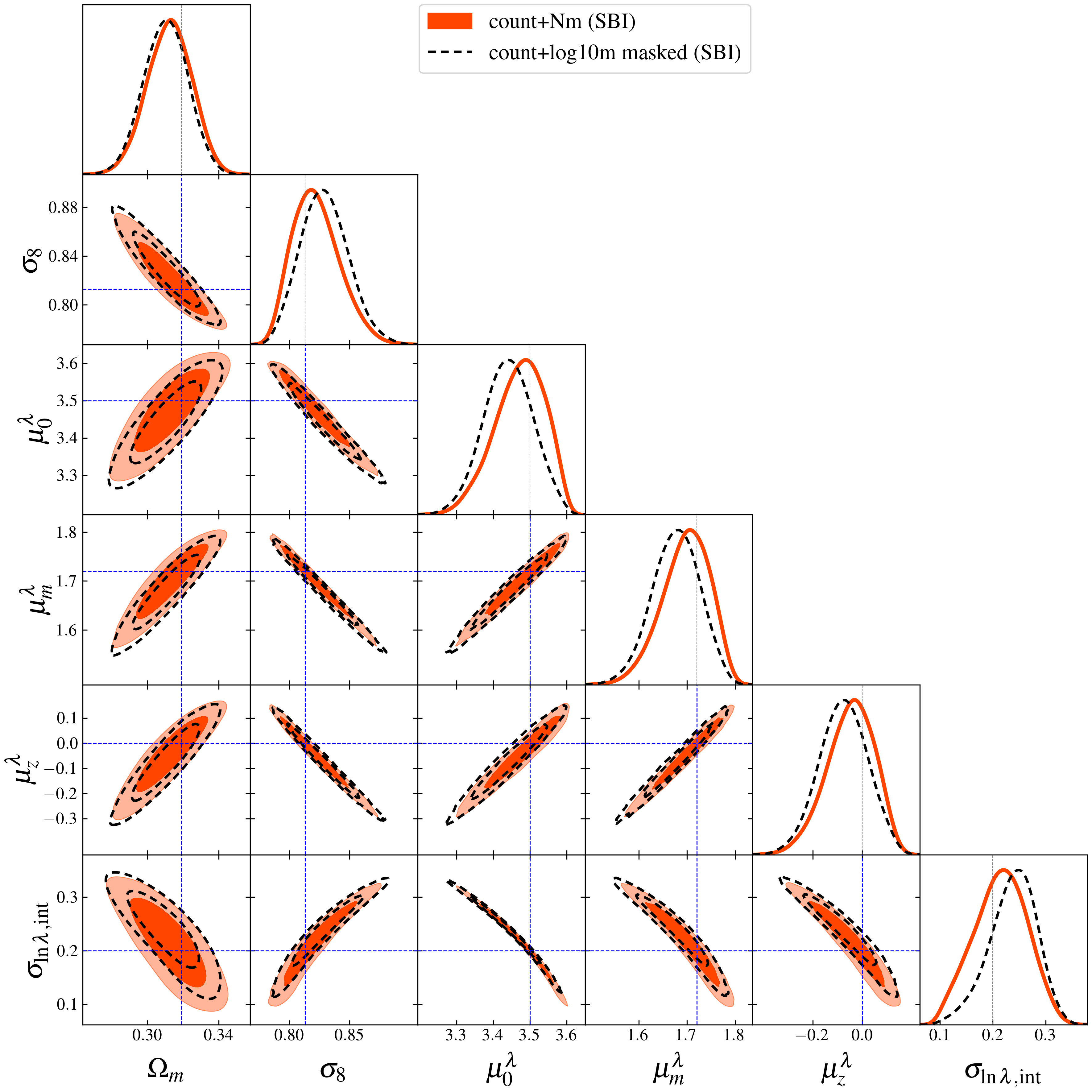}
    \caption{Posterior distributions obtained from the trained posterior generator for the three configurations \texttt{count}, \texttt{Nm}, and \texttt{count\_Nm}, but masking sampled parameters which generated simulations with at least one empty bin.}
    \label{fig:mcmc_masked}
\end{figure}

This selective removal affects the posterior training by effectively imposing a smooth prior in the low-$\Omega_m$–low-$\sigma_8$ region. In Fig. \ref{fig:mcmc_masked} we show the posterior distributions obtained for the idealized data vector using three configurations: \texttt{count\_masked}, \texttt{log10m}, and \texttt{count\_log10m}, where the counts are masked, and the mean logarithmic mass can therefore be computed. Compared to the unmasked case using the appropriate \texttt{Nm} statistics, the resulting posteriors exhibit a small but noticeable shift.

\section{Probability coverage test}
\label{app:restricted_center_probability_coverage}
\begin{figure}
\centering\includegraphics[width=0.4\textwidth]{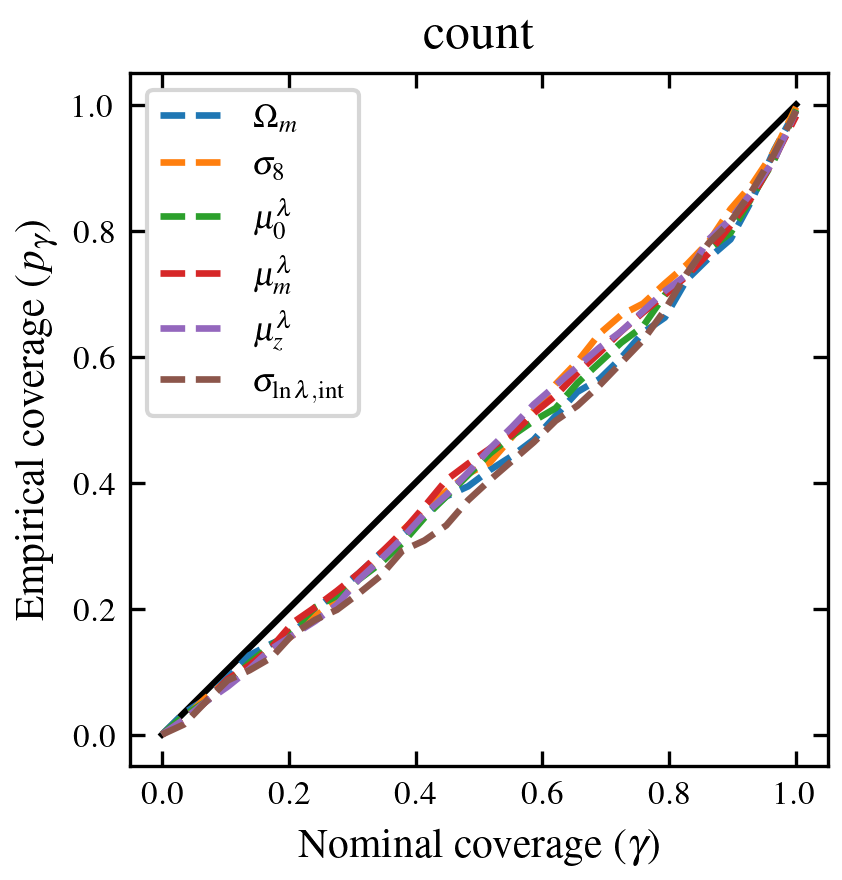}
\includegraphics[width=0.4\textwidth]{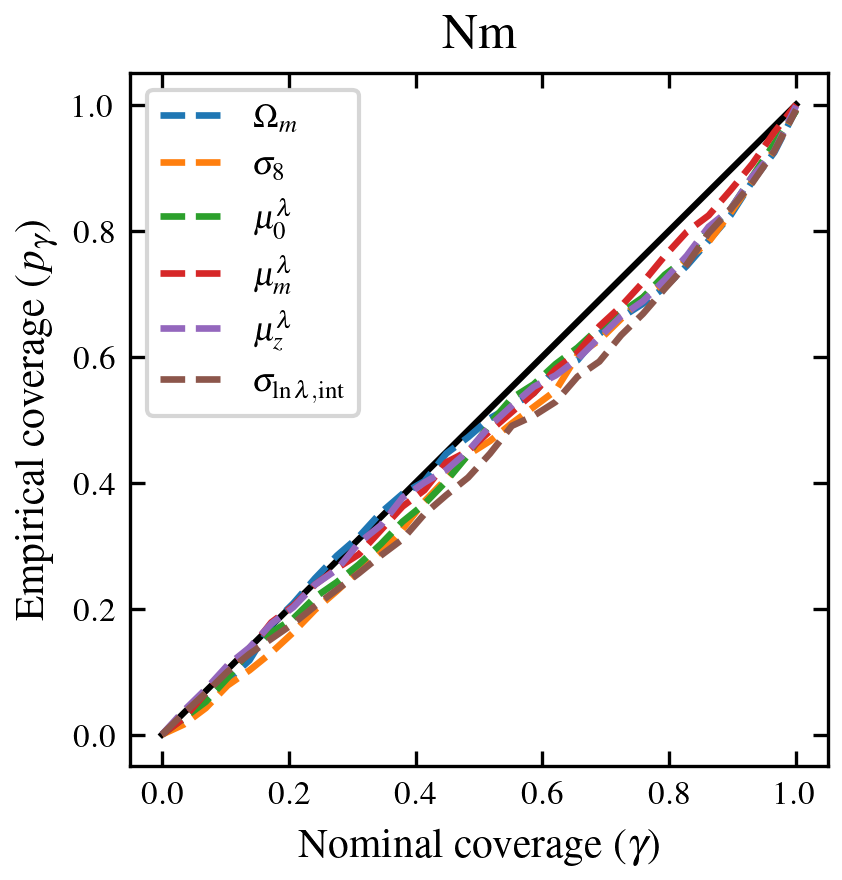}
\includegraphics[width=0.4\textwidth]{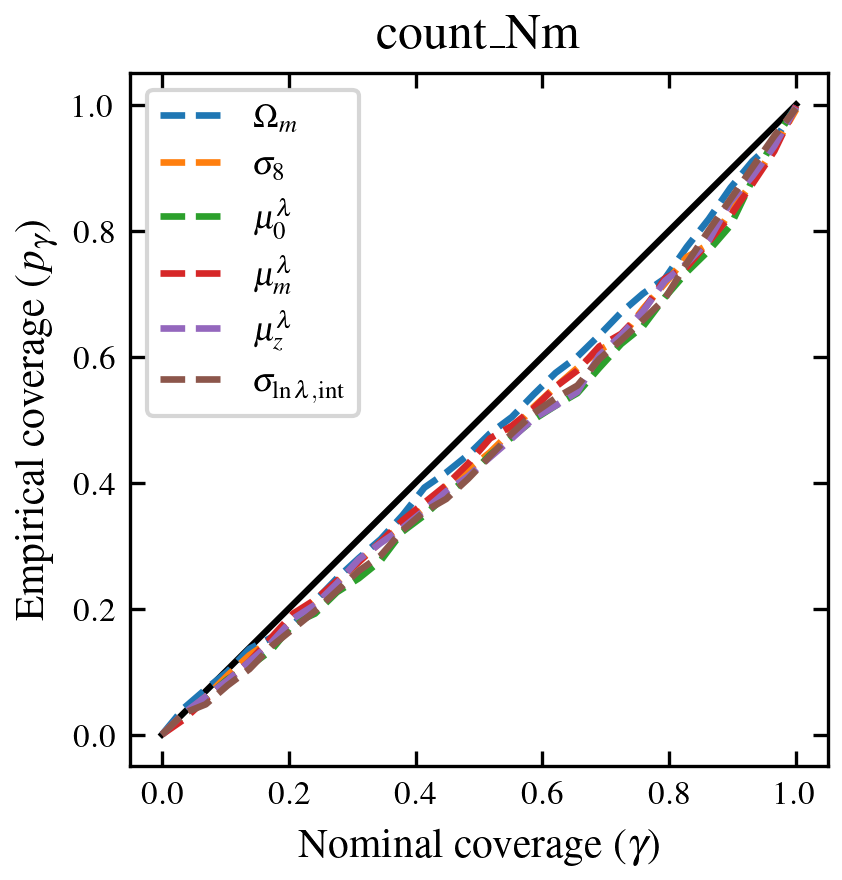}
    \caption{ECP computed for the three posterior estimates. \textit{Top}: Trained on counts only (\texttt{count}); middle: trained on the product of counts and mean masses (\texttt{Nm}). \textit{Bottom}: Trained on the combination of counts and mean masses (\texttt{count\_Nm}). Each panel shows the fraction of true parameter values falling within the corresponding credible intervals, illustrating the calibration of the SBI posteriors.}
    \label{fig:prob_coverage}
\end{figure}

\begin{figure*}
    \centering
\includegraphics[width=0.4\linewidth]{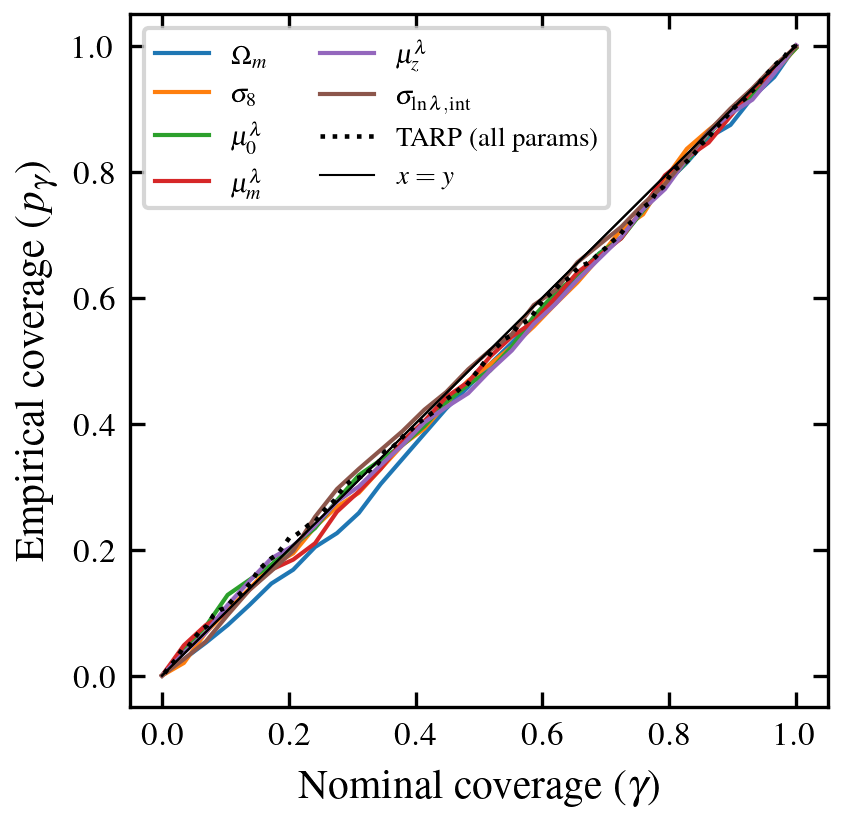}
\includegraphics[width=0.55\linewidth]{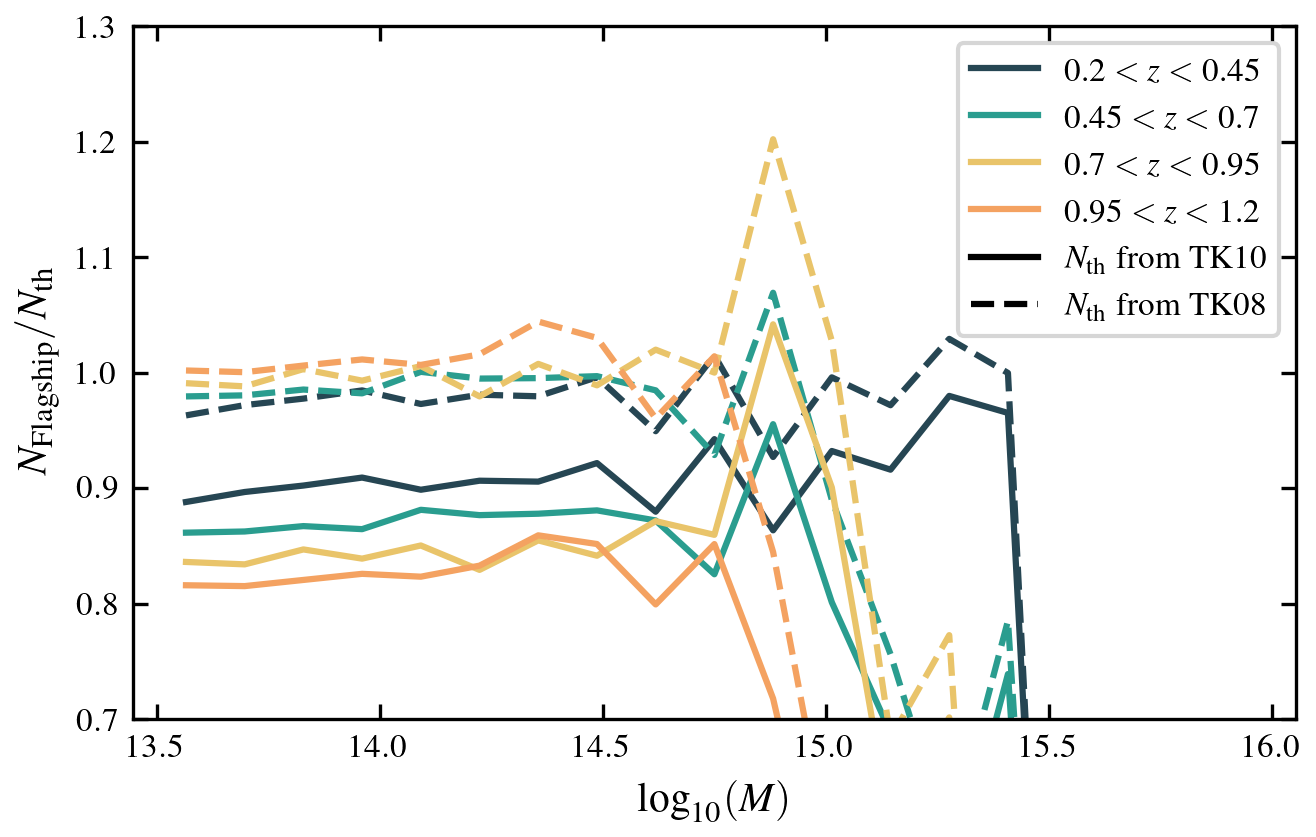}
    \caption{\textit{Left}: Probability coverage tests using the estimated posterior over 6 \capish parameters, given the observation of counts and masses (\texttt{count\_Nm}). To prevent inaccuracies in the density estimation near prior boundaries, the neural posterior is trained on a broad prior, then restricted to a tighter one. The ECP calibration is computed for each parameter, and the TARP calibration is computed for all parameters at once. \textit{Right}: Halo abundance measured in the \textit{Euclid} Flagship simulation, compared to halo abundance prediction based on \citet{Tinker_2010,Tinker2008hmf}.}
    \label{fig:coverage_plot_restricted}
\end{figure*}

As discussed in Sect. \ref{sec:bayesian_robustness}, Bayesian robustness tests are crucial to assess the performance and reliability of NDE for cosmological parameter inference. For each of the 60{,}000 simulations drawn from the prior defined in Sect. \ref{sec:bayesian_robustness}, we draw 5000 samples from the corresponding posterior estimates to obtain a $\gamma$-credible set, for $\gamma \in [0,1]$. The fraction of sets that contain the true parameter converges to the expected coverage probability (ECP) $p_\gamma$, whose calibration curves are shown in Fig. \ref{fig:prob_coverage}. We observe a tendency toward overconfidence (empirical coverage $p_\gamma$ lower than nominal coverage $\gamma$). This is an effect of NDE training when the posterior estimation is performed close to the edge of the parameter prior. This problem is not seen when the posterior estimation is performed away from the prior edge. One solution is to train the NDE within large priors, and evaluate the ECP with the posterior estimates within smaller priors (as mentioned in \ref{sec:bayesian_robustness}), giving Fig. \ref{fig:coverage_plot_restricted} (left panel, shown for the joint \texttt{count\_Nm} setup).

\section{Recovered parameters versus inputs}
\label{app:mean_vs_inputs}
\begin{figure*}
    \centering\includegraphics[width=0.9\textwidth]{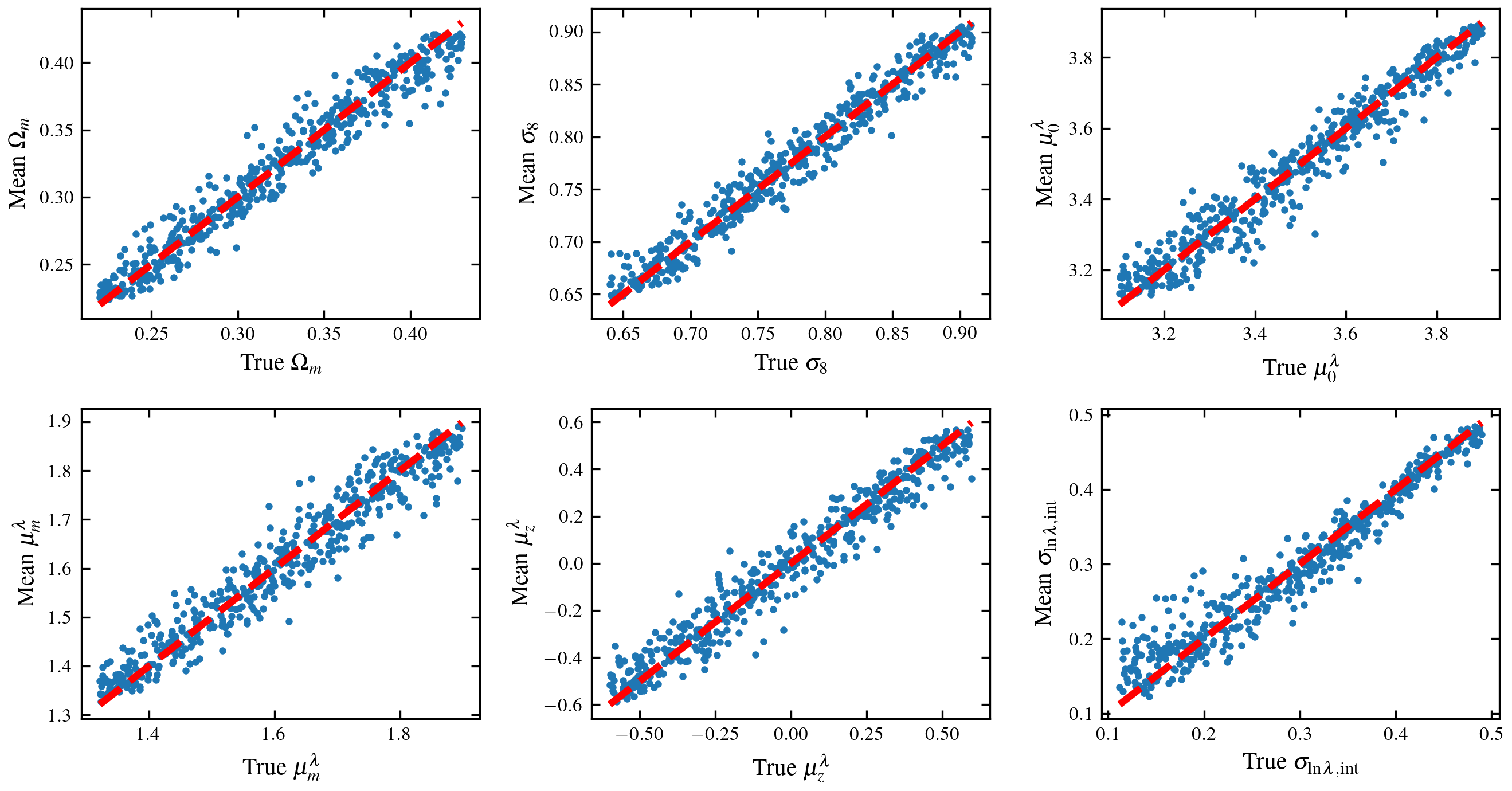}
    \caption{Recovered mean parameters versus true input values for the \texttt{count\_Nm} configuration, computed from 500 simulated pairs $\{\widehat{\theta}_{\mathrm{true},k}, \widehat{D}_k\}$. The red lines indicate the ideal one-to-one relation $x=y$.}
    \label{fig:parameter_bias}
\end{figure*}

Figure \ref{fig:parameter_bias} compares the recovered parameters computed as the mean of \capish posteriors ($y$-axis) versus their input values ($x$-axis).

\section{Analytical likelihood approach}
\label{app:likelihood_standard}
\subsection{Count likelihood}
The analytical prediction for the cluster number count is given by \citep{Payerne2023}
\begin{equation}
\begin{split}
    N_{ij} = \Omega_S\int_{\lambda_i}^{\lambda_{i+1}}\int_{z_j}^{z_{j+1}}&\int_{m_{\rm min}}^{m_{\rm max}}  dz\,dm\,d\lambda_{\rm obs}\\ \,&P(\lambda_{\rm obs}|m,z)\, n_h(m,z).\label{eq:nth_richness_redshift_bins}
    \end{split}
\end{equation}
In the above equation, we consider the mass-richness relation $\ln \lambda_{\rm obs}\sim \mathcal{N}(\langle \ln \lambda_{\rm obs}|m, z \rangle, \sigma_{\ln \lambda}^2)$,  $\sigma_{\ln \lambda}^2$ where $\langle \ln \lambda_{\rm obs} | m, z \rangle$ is given in \eqreff{eq:powerlaw_richness} and $\sigma_{\ln \lambda}^2$ is given in \eqreff{eq:sigma2_lnlambda}, no selection function and no photometric redshift, i.e., $z_{\rm true} = z_{\rm phot}$. We use the  Gaussian likelihood given by
\begin{equation}
    \mathcal{L}_{\rm BLC}^{\rm Gauss-SN+SSC} \propto |\Sigma_N|^{-2}\exp \left[ -\frac{1}{2}[N-\widehat{N}]^T\Sigma_N^{-1}[N-\widehat{N}] \right]
    \label{eq:binned_gaussian_likelihood_count}
\end{equation}
where \begin{equation}
    (\Sigma_N)_{ij,kl} = N_{ij}\, \delta^K_{ik}\, \delta^K_{jl}+ N_{ij}\, N_{kl}\, \langle b\rangle_{ij}\, \langle b\rangle_{kl}\, S_{jl}, 
    \label{eq:covariance}
\end{equation}
is the binned count covariance \citep{Lacasa2023SSC}. The first term is the Poisson shot noise, and the second is the binned SSC contribution, and $\langle b\rangle_{ij}$ is the average halo bias in the given richness-redshift bin.

\subsection{Mass likelihood}

For the mean weak-lensing mass, we consider the Gaussian likelihood
\begin{equation}
    \mathcal{L}_{M_{\rm WL}} \propto |\Sigma_M|^{-2}\,
    \exp\left[-\frac{1}{2}\sum_{ij}
    \left(
    \frac{\log_{10}(M_{ij}) - \log_{10}(\widehat{M}_{ij})}
    {\sigma_{\log_{10}M_{ij}}}
    \right)^2\right],
    \label{eq:binned_gaussian_likelihood_mass}
\end{equation}
where $\widehat{M}_{ij}$ is the observed mean mass in richness--redshift bin $(i,j)$ and $\Sigma_{\log_{10}M}$ is the diagonal covariance of $\log_{10} M$ uncertainties. The theoretical mean mass in each bin is
\begin{equation}
\begin{split}
    M_{ij}^\Gamma = \frac{\Omega_S}{\bar{N}_{ij}}
\int\limits_{\lambda_i}^{\lambda_{i+1}}\int\limits_{z_j}^{z_{j+1}}
    &\int\limits_{m_{\rm min}}^{m_{\rm max}}
    m^\Gamma \, W(z)\, \times\\
&P(\lambda_{\rm obs}|m,z)\, n_h(m,z)\,
    dm\,d\lambda_{\rm obs}\,dz,
\label{eq:mth_richness_redshift_bins}
\end{split}
\end{equation}
where $\bar{N}_{ij}$ is identical to $N_{ij}$ in Eq.~\eqref{eq:nth_richness_redshift_bins} with the addition of the window function $W(z)$. The uncertainty on the mean logarithmic mass is propagated from the Fisher information of the weak-lensing profile for a single cluster:
\begin{equation}
    \sigma_{\log_{10}M_{ij}}(M_{ij}) =
    \frac{\sigma_{M}(M_{ij})}{\ln(10)\,M_{ij}} \frac{1}{\sqrt{N_{ij}}}.
\end{equation}
The uncertainty on the cluster mass $\sigma_M$ is obtained from the inverse Fisher matrix constructed for the parameters
$\boldsymbol{\theta} = (M, c, A)$, where $c$ is the NFW concentration and $A$ is a global multiplicative amplitude factor. The Fisher matrix is defined as
\begin{equation}
    F_{\alpha\beta} =
    \int_{R_{\rm min}}^{R_{\rm max}}
    \frac{1}{\sigma^2_{\Delta\Sigma}(R)}
    \frac{\partial \Delta\Sigma(R)}{\partial \theta_\alpha}
    \frac{\partial \Delta\Sigma(R)}{\partial \theta_\beta}\, dR,
    \label{eq:fisher_mass}
\end{equation}
with $\Delta\Sigma(R)=A\times \Delta\Sigma_{\rm NFW}(R|m,c)$ the predicted excess surface density for an NFW halo of mass $M$, concentration $c$, and amplitude $A$. The amplitude $A$ denotes the possible mis-calibration of source photometric redshifts and shape measurement error. The per-radius dispersion of the excess surface density profile $\sigma^2_{\Delta\Sigma}(R)$ can be modeled as in \eqreff{eq:shape_noise_dispersion}. 
Gaussian priors with width $\sigma_\alpha^2$ can be used to constrain better the systematics like $F_{\alpha\alpha} \rightarrow F_{\alpha\alpha} + 1/\sigma_\alpha^2$. The covariance matrix is then $C_{\alpha\beta} = (F^{-1})_{\alpha\beta}$ and $\sigma_M = \sqrt{C_{MM}}$. 
In this work, we put priors of zero width for the amplitude $A$ and for the concentration $c$, to mimic the error model for individual lensing masses in Appendix \ref{app:error_model_leisng_mass} (which uses a fixed concentration-mass relation, and $A=1$).

\section{Flagship underlying halo mass function}
\label{app:flagship_comparison}
We compute the halo abundance (as described in Sect. \ref{sec:capish_dm_halos}) and compare it with the abundance of dark matter halos identified in the \textit{Euclid} Flagship simulation. The measured abundance is compared to predictions from the halo mass function implementations of \citet{Tinker_2010} and \citet{Tinker2008hmf}. As shown in Fig. \ref{fig:coverage_plot_restricted} (right panel), the bias between the simulated and predicted abundances highlights—consistent with \citet{Euclid2025flagship}—that the \citet{Tinker2008hmf} formulation provides a better match to the simulated dataset.
\section{Cosmological and scaling relation parameter fits}
\begin{table}[h]
\begin{center}
\caption{Means and $1\sigma$ dispersions of parameter posteriors for $\Omega_m$, $\sigma_8$, $\mu_0^\lambda$, $\mu_m^\lambda$, $\mu_z^\lambda$, $\sigma_{\ln \lambda, {\rm int}}$ obtained in this work. }
\label{tab:best_fits}
\resizebox{1\textwidth}{!}{%
\begin{tabular}{ ccccccc } 
  Parameters & $\Omega_m$ & $\sigma_8$ & $\mu_0^\lambda$ & $\mu_m^\lambda$ & $\mu_z^\lambda$ & $\sigma_{\ln \lambda, {\rm int}}$\\
  \hline
\hline
  Fiducial &0.319 & 0.813  & 3.5 & 1.72 & 0 & 0.2\\
  \hline
count (SBI) & $0.343 \pm 0.058$  & $0.845 \pm 0.055$  & $3.396 \pm 0.174$  & $1.704 \pm 0.078$  & $-0.029 \pm 0.176$  & $0.214 \pm 0.053$ \\
Nm (SBI) & $0.314 \pm 0.029$  & $0.804 \pm 0.068$  & $3.533 \pm 0.255$  & $1.649 \pm 0.136$  & $-0.094 \pm 0.221$  & $0.243 \pm 0.077$ \\
count+Nm (SBI) & $0.312 \pm 0.012$  & $0.823 \pm 0.019$  & $3.472 \pm 0.066$  & $1.697 \pm 0.048$  & $-0.049 \pm 0.093$  & $0.215 \pm 0.049$ \\
count-masked (SBI) & $0.330 \pm 0.061$  & $0.842 \pm 0.062$  & $3.417 \pm 0.172$  & $1.694 \pm 0.099$  & $-0.049 \pm 0.219$  & $0.217 \pm 0.054$ \\
log10m (SBI) & $0.330 \pm 0.059$  & $0.871 \pm 0.041$  & $3.530 \pm 0.061$  & $1.732 \pm 0.047$  & $0.055 \pm 0.083$  & $0.197 \pm 0.048$ \\
count+log10m masked (SBI) & $0.310 \pm 0.012$  & $0.829 \pm 0.019$  & $3.441 \pm 0.065$  & $1.677 \pm 0.047$  & $-0.077 \pm 0.094$  & $0.237 \pm 0.044$ \\
count+Nm with selection bias fixed (SBI) & $0.317 \pm 0.011$  & $0.816 \pm 0.015$  & $3.491 \pm 0.055$  & $1.713 \pm 0.040$  & $-0.013 \pm 0.075$  & $0.202 \pm 0.043$ \\
count+Nm with free selection bias (SBI) & $0.317 \pm 0.011$  & $0.817 \pm 0.019$  & $3.489 \pm 0.062$  & $1.708 \pm 0.043$  & $-0.017 \pm 0.084$  & $0.203 \pm 0.045$ \\
count+Nm (SBI, one sim.) & $0.320 \pm 0.013$  & $0.822 \pm 0.020$  & $3.451 \pm 0.071$  & $1.701 \pm 0.051$  & $-0.058 \pm 0.098$  & $0.231 \pm 0.049$ \\
count+Nm (SBI:TK10, Flagship) & $0.307 \pm 0.013$  & $0.807 \pm 0.019$  & $3.482 \pm 0.069$  & $1.737 \pm 0.050$  & $-0.078 \pm 0.099$  & $0.206 \pm 0.052$ \\
count+Nm (SBI:TK08, Flagship) & $0.308 \pm 0.015$  & $0.831 \pm 0.025$  & $3.434 \pm 0.083$  & $1.670 \pm 0.062$  & $-0.084 \pm 0.122$  & $0.240 \pm 0.055$ \\

  \hline
count (analytic likelihood) & $0.327 \pm 0.048$  & $0.824 \pm 0.048$  & $3.459 \pm 0.141$  & $1.710 \pm 0.083$  & $-0.012 \pm 0.180$  & $0.212 \pm 0.058$ \\
log10m (analytic likelihood) & $0.320 \pm 0.072$  & $0.823 \pm 0.055$  & $3.496 \pm 0.057$  & $1.721 \pm 0.044$  & $-0.002 \pm 0.082$  & $0.204 \pm 0.047$ \\
count+log10m (analytic likelihood) & $0.316 \pm 0.008$  & $0.816 \pm 0.012$  & $3.490 \pm 0.044$  & $1.713 \pm 0.032$  & $-0.012 \pm 0.062$  & $0.206 \pm 0.033$ \\
\end{tabular}}
\tablefoot{The first block lists the fiducial values used in this work, the second block lists the constraints on the same parameters obtained with SBI, the third block lists analytic-likelihood constraints.}
\end{center}
\end{table}
We recap in \tabreff{tab:best_fits} the different constraints on the cosmological parameters $\Omega_m$ and $\sigma_8$, as well as on the scaling-relation parameters, obtained in this work. The first line corresponds to the fiducial values, the second block to the SBI results, and the final block to the analytic likelihood results.
\end{appendix}
\end{document}